\documentclass[pdflatex,sn-mathphys-num]{sn-jnl}

\usepackage{amsmath,amssymb,amsfonts}
\usepackage[T1]{fontenc}
\usepackage{booktabs}
\usepackage{adjustbox}

\usepackage{subcaption}
\usepackage{comment}
\usepackage{tikz}
\usepackage{placeins}
\usepackage{pdfpages}
\usepackage{caption}
\usepackage{hyperref}
\usepackage{cleveref}
\usepackage{multirow}%
\usepackage{amsthm}%
\usepackage{mathrsfs}%
\usepackage[title]{appendix}%
\usepackage{xcolor}
\usepackage{textcomp}%
\usepackage{manyfoot}%
\usepackage{algorithm}%
\usepackage{algorithmicx}%
\usepackage{algpseudocode}%
\usepackage{listings}%

\DeclareCaptionLabelSeparator{line}{\textbf{|}}

\usepackage{float}   
\newcommand{\figlabelsize}{\normalsize}

\usepackage{graphicx}
\graphicspath{{figures/}}

\begin{document}

\title{Persistent geographical biases in global scientific collaboration and citations}


\author*[1,2]{\fnm{Leyan} \sur{Wu}}\email{wuleyan@whu.edu.cn}

\author[2,3]{\fnm{Yong} \sur{Huang}}\email{yonghuang1991@whu.edu.cn}

\author[2,3]{\fnm{Wei} \sur{Lu}}\email{weilu@whu.edu.cn}

\author[4]{\fnm{Akrati} \sur{Saxena}}\email{a.saxena@liacs.leidenuniv.nl}

\author[5]{\fnm{Vincent} \sur{Traag}}\email{v.a.traag@cwts.leidenuniv.nl}


\affil*[1]{\orgdiv{School of Political Science and Public Administration},\orgname{Wuhan University}, 
\orgaddress{\city{Wuhan}, \postcode{430072}, \country{China}}}

\affil[2]{\orgdiv{Institute of Intelligence and Innovation Governance},\orgname{Wuhan University}, 
\orgaddress{\city{Wuhan}, \postcode{430072}, \country{China}}}

\affil[3]{\orgdiv{School of Information Management},\orgname{Wuhan University}, 
\orgaddress{\city{Wuhan}, \postcode{430072}, \country{China}}}

\affil[4]{\orgdiv{Leiden Institute of Advanced Computer Science (LIACS)},
\orgname{Leiden University},
\orgaddress{\country{The Netherlands}}}

\affil[5]{\orgdiv{Centre for Science and Technology Studies (CWTS)},
\orgname{Leiden University},
\orgaddress{\country{The Netherlands}}}

\abstract{Scientific knowledge flows enable cumulative progress by connecting researchers across disciplines, institutions, and countries. Yet it remains unclear how geography and national structures continue to shape these exchanges in an increasingly connected world. Using a large-scale bibliometric dataset from OpenAlex, which covers 39.35 million publications across 95 countries and 3,794 cities between 2000 and 2022, we examine global knowledge diffusion through two complementary channels: co-authorship and citation. 
We find that the constraining effect of geographic distance on collaboration has not diminished over time but has instead intensified, suggesting persistent structural or institutional barriers. Citation flows, by contrast, are less sensitive to spatial proximity, indicating that intellectual influence may diffuse more freely across borders.
At the country level, research networks exhibit strong domestic preferences and a shared citation orientation toward the United States. China, while increasingly favored as a collaboration partner by other countries, continues to be systematically undercited within global citation flows. International mobility increases researchers' collaboration with scholars in their host country but has limited effects on citation flows.
These results highlight the structural persistence of spatial and country biases in global science, with implications for equitable participation and recognition across regions.}


\maketitle              

\section{Introduction}

Knowledge flow is widely regarded as the foundation of scientific progress. The exchange of ideas, methods and discoveries enables cumulative progress, connecting researchers across fields, institutions and borders~\cite{zhuge2006discovery,megnigbeto2015effect,frenken2020geography,wang2021knowledge}. In principle, the scientific enterprise aspires to be a borderless system where knowledge circulates freely, unbounded by geography~\cite{cairncross_death_1997,frenken2019death}. Whether this ideal of global openness is realized in practice remains uncertain. The movement of knowledge across space determines not only the pace of science but also who participates in and benefits from science, shaping visibility, recognition and opportunity within the global scientific community~\cite{nettasinghe2021emergence,miao2024persistent}. Therefore, examining the actual patterns, drivers and barriers of global knowledge flows is crucial for addressing the challenges of international scientific collaboration and for building a more equitable and efficient global research ecosystem.

Knowledge flows in science can be studied through the connections between researchers and their work. Specifically, empirical studies commonly consider collaborations and citations as indicators of knowledge flows~\cite{jaffe1993geographic,singh2005collaborative,sorenson2007science,sidone2017scholarly}. Collaboration captures the direct exchange of ideas and joint production of new knowledge~\cite{nelson2009measuring,capello2018proximities,newman2004coauthorship}, whereas citation reflects the intellectual lineage through which discoveries diffuse across the global literature~\cite{yan2012scholarly}. Together, these two dimensions offer complementary perspectives on how knowledge travels across spatial and institutional boundaries. 
Previous studies have shown that linguistic, cultural, and institutional proximity play a crucial role in fostering cross-border scientific connections~\cite{bassecoulard2000insights,okubo2004searching,leonard2012cooperation}. The influence of geographic proximity, however, remains a central topic of debate, also in the digital era. While advances in communication technologies and virtual collaboration tools have expanded opportunities for long-distance interaction, empirical studies still find that collaboration probabilities decline with geographic distance \cite{isfandyari2023global,pan2012world,van2023impact,toobaee2024proximity}. National borders may similarly continue to shape scientific activity. Researchers have observed a tendency to cite domestic authors~\cite{lariviere2018citations}, and studies indicate that research originating from some countries receives fewer citations while other countries receive more citations~\cite{meneghini2008articles,akre2011differences}. These patterns suggest that country-specific biases may influence scholarly communication~\cite{paris1998region,gomez2022leading}. Concerns over the unequal distribution of global citation flows have further highlighted issues of academic inequality~\cite{gomez2022leading,nielsen2021global}. These findings suggest that scientific visibility and collaboration are not evenly distributed, but deeply shaped by spatial and geopolitical factors.

However, the interplay between geography and scientific practice, both within and between countries, remains insufficiently understood. Some earlier findings presented as evidence of a country bias may actually reflect more subtle geographic effects. Geographic distance can influence not only international collaboration and citation but also domestic scientific interactions, particularly in countries where research infrastructure is unevenly developed across regions. This ambiguity highlights the limitations of relying solely on country-level analyses. It also underscores the need for a more detailed examination of how geography shapes patterns of collaboration and citation at multiple scales.

\begin{figure}
    \centering
    \includegraphics[width=0.5\linewidth]{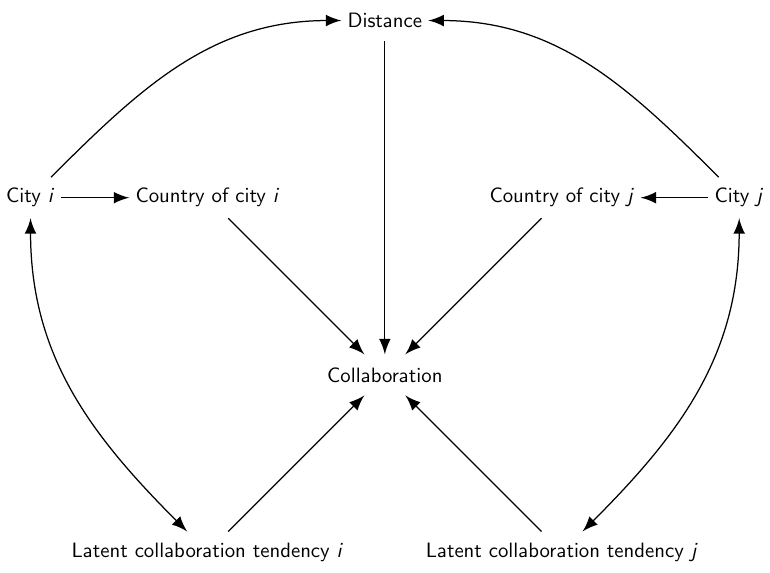}
    \caption{A simple directed acyclic graph (DAG) to illustrate the core causal assumptions to identify potential biases of the country. In particular, we are interested in the direct causal effect of the country on collaborations. Note that the influence of scientific fields is not illustrated in this DAG, but it is assumed to act simply as a confounder between the city and the collaboration.  Although this DAG only illustrates the assumptions for collaborations, a DAG for citations would be highly similar, except that time plays an additional complication.}
    \label{fig:dag_collaboration}
\end{figure}

Moreover, identifying and even defining what we mean by a bias is not straightforward.
We here rely on the terminology introduced by \cite{traag_causal_2022} and understand a bias as an unjustified direct causal effect.
We illustrate our core causal assumptions in order to identify such a bias in the context of collaborations in Fig.~\ref{fig:dag_collaboration}.
As shown previously~\cite{pan2012world}, collaboration is likely to be affected by the distance.
If we ignore this effect, we are not identifying a direct causal effect of the country on collaboration, because of the confounding influence through the city and the distance.
Hence, we need to close this non-causal path by explicitly controlling for distance.
Moreover, there are latent tendencies to collaborate with a certain city.
This latent collaboration tendency can represent a multitude of factors, encompassing various factors such as scientific quality and strength, but also reputational factors, such as well-known authors or institutions that attract more collaborations.
We consider such a latent collaboration tendency and also control for this in our analysis.

Building on this foundation, we address a central question: How do geographic distance and country-level factors shape international scientific collaboration and citation? To answer this question, we draw on OpenAlex, a comprehensive academic database, to construct a large-scale bibliometric dataset that links scientific publications to the institutional locations of their authors. The dataset covers 39.35 million publications from 2000 to 2022, spanning 95 countries and 3,794 cities, and includes 2,444,439 city-level collaboration pairs and 14,754,699 city-level citation pairs \footnote{The difference in counts arises from fundamental distinctions: collaboration pairs are undirected (each unique city pair counted once), while citation pairs are directed (A→B and B→A count as two separate pairs). Additionally, citation relationships are inherently more frequent than co-authorship collaborations, and the cumulative nature over 23 years amplifies this disparity.}, providing a detailed perspective on global research dynamics. We employ a Bayesian modeling framework to estimate the influence of geographic distance on cross-country collaboration and citation patterns, and to examine bilateral preferences in collaboration and citation between countries. In addition, we apply a matching design to study how international mobility affects researchers' subsequent collaboration networks and citation visibility. Detailed descriptions of data cleaning, geographic labeling, and modeling procedures are provided in the \nameref{sec:methods}. By systematically analyzing the influence of spatial and national structures on scientific interactions, this study advances a comprehensive understanding of global knowledge diffusion, demonstrating how geography and national boundaries continue to shape collaboration, recognition, and collective creativity.

\section{Results}

\subsection{Impact of geographic distance on international collaboration and citation}
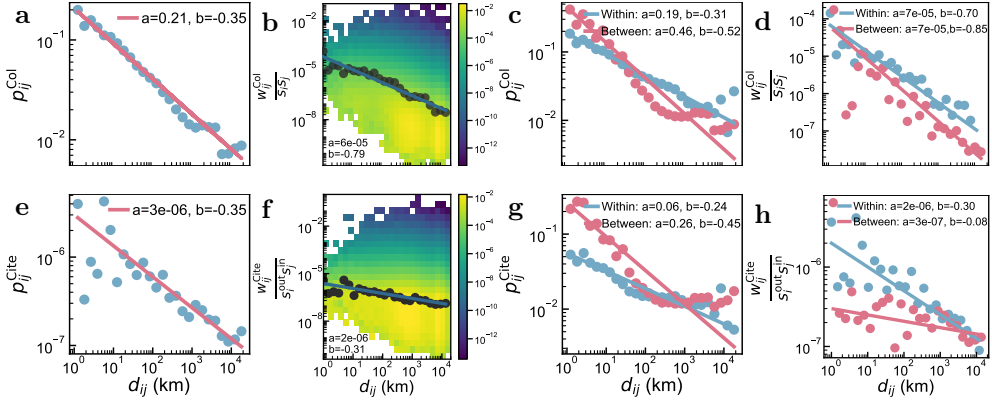
\begin{figure}[htbp]
  \centering
  \begin{tabular}{@{}c@{}c@{}c@{}c@{}}

    \begin{subfigure}{0.248\linewidth}
      \centering
      \begin{tikzpicture}
        \node[anchor=south west, inner sep=0] (img) at (0,0)
          {\includegraphics[width=\linewidth]{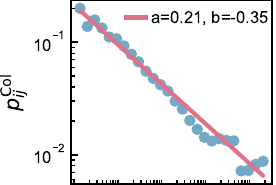}};
        \node[anchor=north west] at (img.north west) {\figlabelsize\textbf{a}};
      \end{tikzpicture}
      \caption{}
      \label{fig:grid-a}
    \end{subfigure} &

    \begin{subfigure}{0.248\linewidth}
      \centering
      \begin{tikzpicture}
        \node[anchor=south west, inner sep=0] (img) at (0,0)
          {\includegraphics[width=\linewidth]{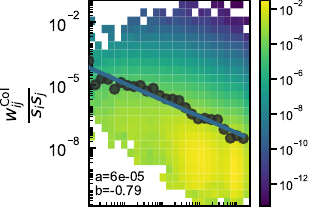}};
        \node[anchor=north west] at (img.north west) {\figlabelsize\textbf{b}};
      \end{tikzpicture}
      \caption{}
      \label{fig:grid-b}
    \end{subfigure} &

    \begin{subfigure}{0.248\linewidth}
      \centering
      \begin{tikzpicture}
        \node[anchor=south west, inner sep=0] (img) at (0,0)
          {\includegraphics[width=\linewidth]{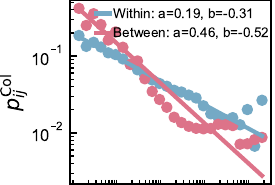}};
        \node[anchor=north west] at (img.north west) {\figlabelsize\textbf{c}};
      \end{tikzpicture}
      \caption{}
      \label{fig:grid-c}
    \end{subfigure} &

    \begin{subfigure}{0.248\linewidth}
      \centering
      \begin{tikzpicture}
        \node[anchor=south west, inner sep=0] (img) at (0,0)
          {\includegraphics[width=\linewidth]{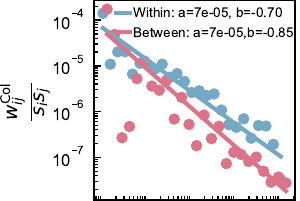}};
        \node[anchor=north west] at (img.north west) {\figlabelsize\textbf{d}};
      \end{tikzpicture}
      \caption{}
      \label{fig:grid-d}
    \end{subfigure} \\[-4ex]

    \begin{subfigure}{0.248\linewidth}
      \centering
      \begin{tikzpicture}
        \node[anchor=south west, inner sep=0] (img) at (0,0)
          {\includegraphics[width=\linewidth]{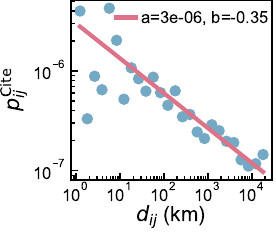}};
        \node[anchor=north west] at (img.north west) {\figlabelsize\textbf{e}};
      \end{tikzpicture}
      \caption{}
      \label{fig:grid-e}
    \end{subfigure} &

    \begin{subfigure}{0.248\linewidth}
      \centering
      \begin{tikzpicture}
        \node[anchor=south west, inner sep=0] (img) at (0,0)
          {\includegraphics[width=\linewidth]{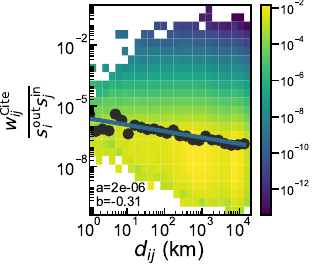}};
        \node[anchor=north west] at (img.north west) {\figlabelsize\textbf{f}};
      \end{tikzpicture}
      \caption{}
      \label{fig:grid-f}
    \end{subfigure} &

    \begin{subfigure}{0.248\linewidth}
      \centering
      \begin{tikzpicture}
        \node[anchor=south west, inner sep=0] (img) at (0,0)
          {\includegraphics[width=\linewidth]{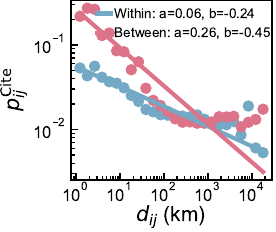}};
        \node[anchor=north west] at (img.north west) {\figlabelsize\textbf{g}};
      \end{tikzpicture}
      \caption{}
      \label{fig:grid-g}
    \end{subfigure} &

    \begin{subfigure}{0.248\linewidth}
      \centering
      \begin{tikzpicture}
        \node[anchor=south west, inner sep=0] (img) at (0,0)
          {\includegraphics[width=\linewidth]{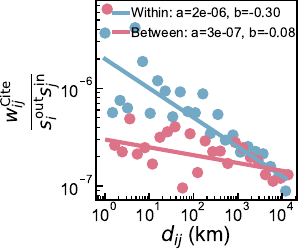}};
        \node[anchor=north west] at (img.north west) {\figlabelsize\textbf{h}};
      \end{tikzpicture}
      \caption{}
      \label{fig:grid-h}
    \end{subfigure}

  \end{tabular}

  \caption{\textbf{Overall trends in collaboration and citation metrics.}
  \textbf{a}, Probability of the existence of a link as a function of the geographic distance between two cities in the collaboration network. 
\textbf{b}, Heatmap shows the distribution of the ratio of the link weight and the product of the strengths of its endpoints $ w_{ij}^{\mathrm{Col}} / (s_i s_j) $ in the collaboration network against the geographic distance $ d_{ij} $ between cities; scatter points and lines represent the average ratio per distance bin.  
\textbf{c}, The trends shown in a, separated by collaboration type into domestic and international. 
\textbf{d}, The trends shown in b, separated by collaboration type into domestic and international. 
\textbf{e}, Probability of the existence of a link as a function of the geographic distance between two cities in the citation network. 
\textbf{f}, Heatmap shows the distribution of the ratio of the link weight and the product of the strengths of its endpoints $ w_{ij}^{\mathrm{Cite}} / (s_i s_j) $ in the citation network against the geographic distance $ d_{ij} $ between cities; scatter points and lines represent the average ratio per distance bin. 
\textbf{g}, The trends shown in e, separated by collaboration type into domestic and international. 
\textbf{h}, The trends shown in f, separated by collaboration type into domestic and international.}
  \label{fig:grid-tightest}
\end{figure}

We first examine the overall influence of geography on collaboration and citation dynamics between cities (Figs.~\ref{fig:grid-a} and \ref{fig:grid-e}). Both the probability of collaboration and citation decline with increasing geographic distance following a power-law pattern, with geographic distance having a stronger effect on collaboration than on citations. A gravity model fits both types of networks well (Figs.~\ref{fig:grid-b} and~\ref{fig:grid-f}). Further analysis reveals that within countries, the probability of collaboration decreases more slowly with distance, while across countries, the decline is steeper. At larger distances, the likelihood of collaboration remains significantly higher within countries compared to between countries under similar distance conditions (Figs.~\ref{fig:grid-c} and~\ref{fig:grid-d}). In terms of citations, the rate within countries shows weaker dependence on distance ($b = -0.24$), whereas cross-country citations decline more clearly with distance ($b = -0.45$) (Fig.~\ref{fig:grid-g}).

\begin{figure}[ht]
    \centering
    \includegraphics[width=0.8\textwidth]{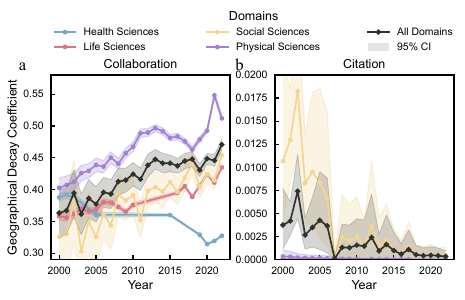} 
    \caption{%
    \textbf{Impact of geographic distance on international collaboration and citation across disciplines.}
    \textbf{a}, Temporal trend of the effect of geographic distance on international collaboration,
    as estimated by our statistical model, controlling for publication volume and country-level preferences.
    The black line represents the average trend across all disciplines, while colored lines denote individual
    fields. \textbf{b}, Analogous results for international citations. The shading around the trends denotes the confidence intervals of the estimated coefficients.}
    \label{fig:2}
\end{figure}

We then analyze the impact of geographic distance on international collaboration and citation patterns using the statistical model, while controlling for both the publication volume of each city and the collaboration preferences of its respective country. We fit the model separately for each field and publication year. The results show that the effect of geographic distance on international collaboration has generally increased over time (the black line in Fig.~\ref{fig:2}a), with the average coefficient across all fields rising from 0.37 in 2000 to 0.48 in 2022. To further explore disciplinary variations, we categorized the data into four domains based on the ``concept'' field of the papers: Physical Sciences, Social Sciences, Life Sciences, and Health Sciences. The mapping between OpenAlex concepts and these four fields is detailed in Supplementary Table S1. When considering disciplines, the impact of geographic distance is most pronounced in the Physical Sciences (the purple line in Fig.~\ref{fig:2}a). Over time, the influence of geographic distance in the Health Sciences has shown a gradual decline, although there has been a slight uptick post-2020 (the blue line in Fig.~\ref{fig:2}a). However, the estimation procedure showed some difficulties in the Health Sciences, and the statistical model may be less appropriate for this domain. In contrast, other disciplines follow the overall trend of increasing impact.

The effect of geographic distance on international citations is much smaller. As shown in Fig.~\ref{fig:2}b, the overall impact of distance on citations is minimal (with a maximum coefficient of approximately 0.02) and has been declining since 2001. From a disciplinary perspective, the Social Sciences experience the greatest impact, which aligns with the overall trend over time. In other domains, the effect is nearly negligible.

\subsection{Statistical modelling}

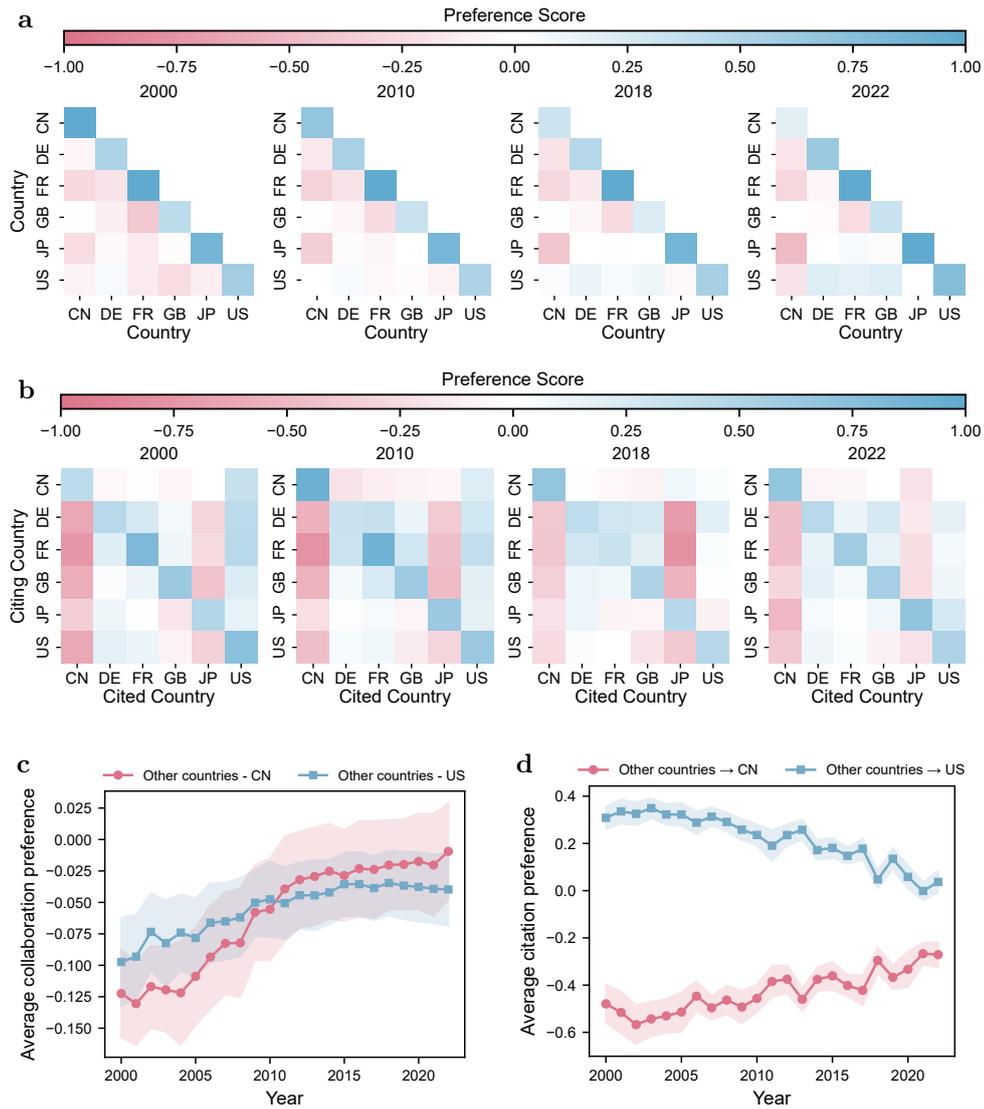
\begin{figure}[!htbp]
    \centering

    \begin{subfigure}[b]{0.98\linewidth}
        \centering
        \begin{tikzpicture}
            \node[inner sep=0] (img)
                {\includegraphics[width=\linewidth]{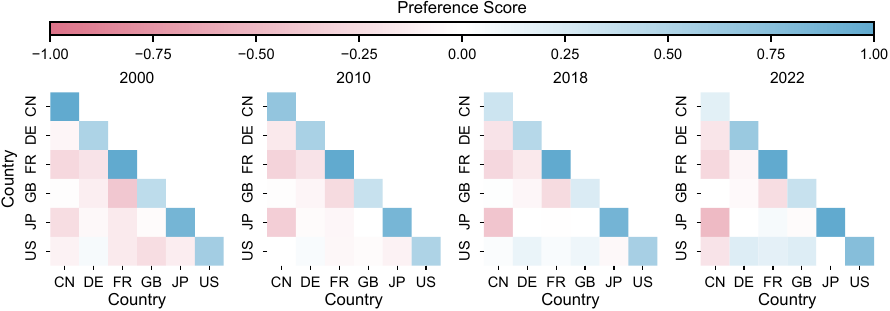}};
            \node[anchor=north west] at (img.north west) {\figlabelsize\textbf{a}};
        \end{tikzpicture}
        \caption{}
        \label{fig:combined_a}
    \end{subfigure}

    \vspace{-14pt}

    \begin{subfigure}[b]{0.98\linewidth}
        \centering
        \begin{tikzpicture}
            \node[inner sep=0] (img)
                {\includegraphics[width=\linewidth]{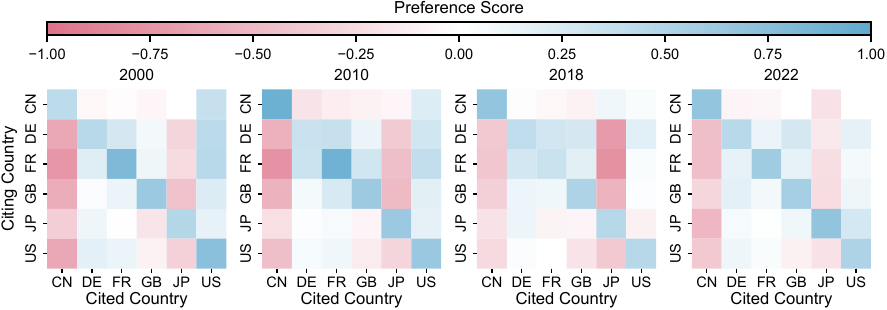}};
            \node[anchor=north west] at (img.north west) {\figlabelsize\textbf{b}};
        \end{tikzpicture}
        \caption{}
        \label{fig:combined_b}
    \end{subfigure}

    \setcounter{subfigure}{2}

    \vspace{-10pt}

    \begin{tabular}{@{}c@{\hspace{3mm}}c@{}}

        \begin{subfigure}{0.48\linewidth}
          \centering
          \begin{tikzpicture}
            \node[anchor=south west, inner sep=0] (img1) at (0,0) 
              {\includegraphics[width=\linewidth]{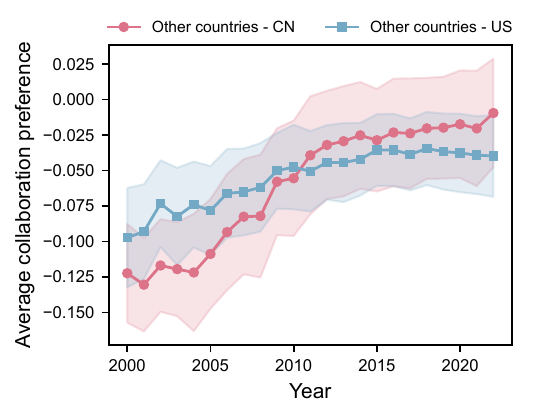}};
            \node[anchor=north west] at (img1.north west) {\figlabelsize\textbf{c}};
          \end{tikzpicture}
          \caption{}
          \label{fig:4c}
        \end{subfigure} &
        
        \begin{subfigure}{0.48\linewidth}
          \centering
          \begin{tikzpicture}
            \node[anchor=south west, inner sep=0] (img2) at (0,0) 
              {\includegraphics[width=\linewidth]{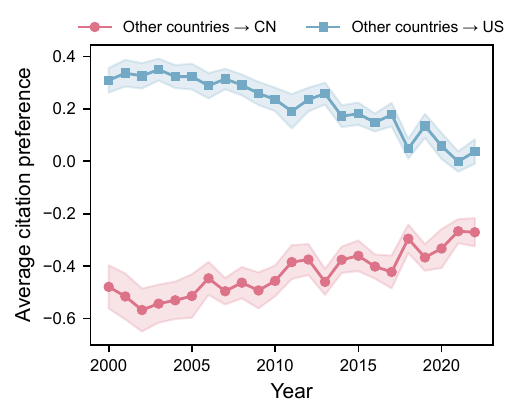}};
            \node[anchor=north west] at (img2.north west) {\figlabelsize\textbf{d}};
          \end{tikzpicture}
          \caption{}
          \label{fig:4d}
        \end{subfigure} \\

    \end{tabular}

    \vspace{-20pt}

    \caption{\textbf{International collaboration and citation preferences between countries.}
    \textbf{a}, Overall collaboration preference among six selected countries from 2000 to 2022.  
    \textbf{b}, Overall citation preference among the same countries and years. 
    \textbf{c}, Annual trends of collaboration preference of all countries toward China and the United States.  
    \textbf{d}, Annual trends of citation preference of all countries toward China and the United States.  
    Both heatmaps display bilateral preferences across all disciplines, where lighter pink indicates lower preference and deeper blue indicates stronger collaboration or citation strength. The shaded areas of the lines represent the 95\% confidence intervals of the means of data in each year.}
    \label{fig:collab_citation_preference}
\end{figure}

Our analysis reveals clear evidence of strong country-specific effects in scientific collaboration (Fig.~\ref{fig:combined_a}). Among the six most research-active countries (US, CN, GB, JP, FR and DE; country abbreviations and their full names are listed in Supplementary Table~S2), collaboration has remained predominantly domestic throughout the years, with a modest tendency to collaborate more frequently with the United States. Collaboration among all countries is substantially more dispersed, exhibiting no consistent pattern of international collaboration preferences (Supplementary Fig.~S1). A similar trend is observed in citation behavior (Fig.~\ref{fig:combined_b}), where most countries predominantly cite domestic research, although the effect is less pronounced than for collaborations. The United States is more likely to be cited by other countries, whereas China and Japan are cited the least. Citation behavior across all countries exhibits largely the same overall pattern (Supplementary Fig.~S2). We further extended our analysis to include all countries to examine how they collaborate with and cite the United States and China (Figs.~\ref{fig:4c},~\ref{fig:4d}). As shown in Fig.~\ref{fig:4c}, before 2010, collaboration with the United States was generally more preferred than with China, but the trend gradually reversed thereafter, a shift that coincides with China’s rapid expansion in research capacity and its rising prominence as an international collaboration partner. In contrast, citation patterns display a pronounced asymmetry: although other countries’ citation preference for China has improved over time, it remains negative overall, whereas their preference for citing work from the United States remains consistently strong and positive. This asymmetric pattern is consistently observed across disciplines (Supplementary Figure~S3).

\begin{figure}[htbp]
  \centering

  \begin{tabular}{@{}c@{\hspace{3mm}}c@{}}

    \begin{subfigure}{0.48\linewidth}
      \centering
      \begin{tikzpicture}
        \node[anchor=south west, inner sep=0] (img1) at (0,0) 
          {\includegraphics[width=\linewidth]{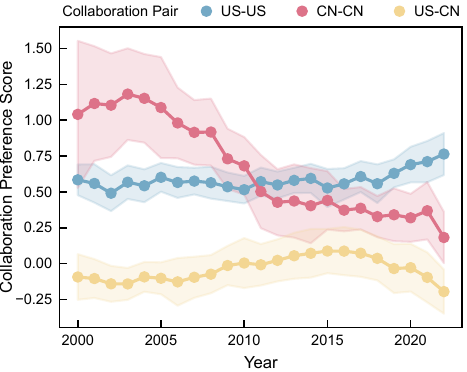}};
        \node[anchor=north west] at (img1.north west) {\figlabelsize\textbf{a}};
      \end{tikzpicture}
      \caption{}
      \label{fig:5a}
    \end{subfigure} &

    \begin{subfigure}{0.48\linewidth}
      \centering
      \begin{tikzpicture}
        \node[anchor=south west, inner sep=0] (img2) at (0,0) 
          {\includegraphics[width=\linewidth]{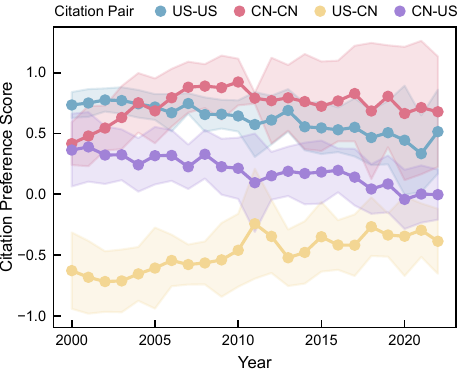}};
        \node[anchor=north west] at (img2.north west) {\figlabelsize\textbf{b}};
      \end{tikzpicture}
      \caption{}
      \label{fig:5b}
    \end{subfigure} \\
    
  \end{tabular}
  \\[-6ex]

  \begin{tabular}{@{}c@{}}
    \begin{subfigure}{0.98\linewidth}
      \centering
      \begin{tikzpicture}[clip=false]
        \node[anchor=south west, inner sep=0] (img3) at (0,0) 
          {\includegraphics[width=\linewidth, trim=0 0 0 -10, clip=false]{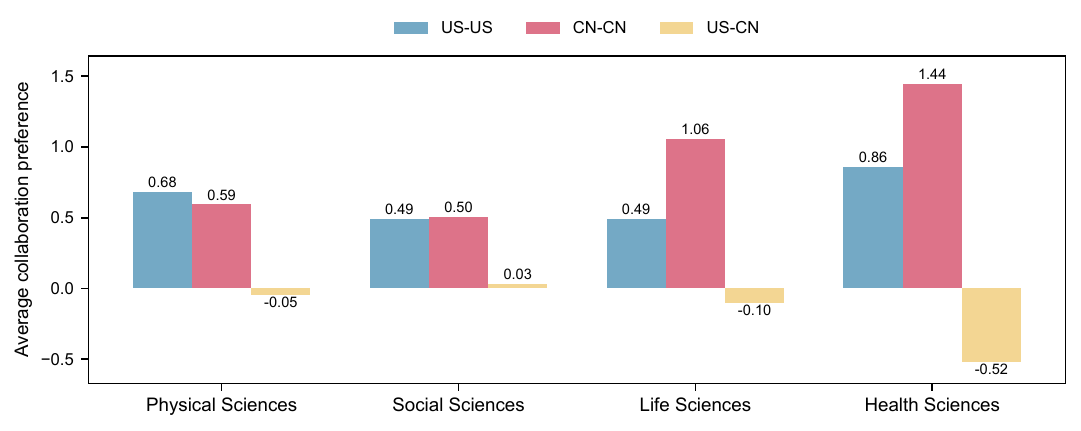}};
        \node[anchor=north west] at (img3.north west) {\figlabelsize\textbf{c}};
      \end{tikzpicture}
      \caption{}
      \label{fig:5c}
    \end{subfigure} \\
  
  \end{tabular}
  \\[-7.5ex]
  
  \begin{tabular}{@{}c@{}}
    \begin{subfigure}{0.98\linewidth}
      \centering
      \begin{tikzpicture}[clip=false]
        \node[anchor=south west, inner sep=0] (img4) at (0,0) 
          {\includegraphics[width=\linewidth, trim=0 0 0 -3, clip=false]{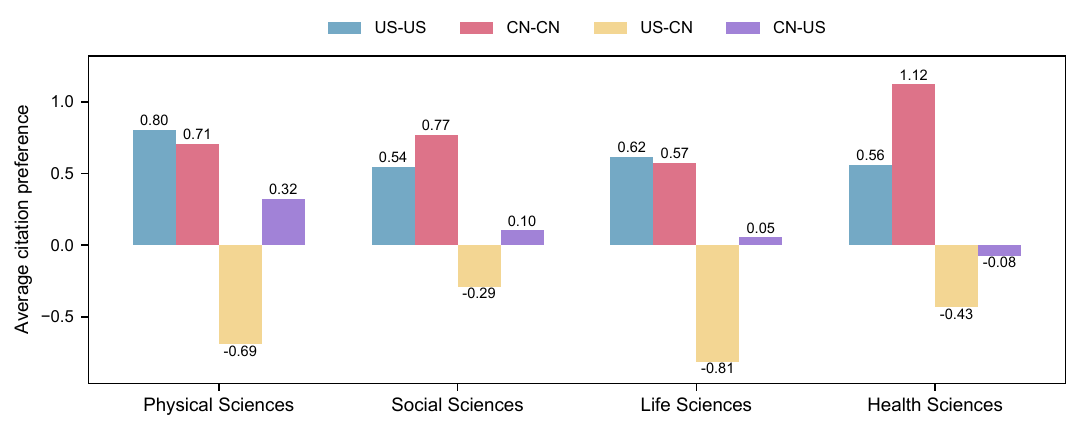}};
        \node[anchor=north west, yshift=10pt] at (img4.north west) {\figlabelsize\textbf{d}};
      \end{tikzpicture}
      \caption{}
      \label{fig:5d}
    \end{subfigure} \\
  \end{tabular}

  \vspace{-18pt}

  \caption{\textbf{Collaboration and citation preference (US-CN).}
  \textbf{a}, The evolution of collaboration preference between the US and China over time. Collaboration preference is undirected. 
  \textbf{b}, The change in citation preference between the two countries over time. Citation preference is directed: "CN-US" refers to Chinese papers citing US papers, while "US-CN" refers to US papers citing Chinese papers.
  Shaded areas in panels a and b represent the 95\% confidence intervals of yearly means. 
  \textbf{c}, Field-specific total collaboration preference, aggregated from panel a.
  \textbf{d}, Field-specific total citation preference, aggregated from panel b.}
  \label{fig:us-cn_preference}
\end{figure}

Furthermore, we examined the collaboration and citation preferences between the United States and China (Fig.~\ref{fig:us-cn_preference}). Fig.~\ref{fig:5a} shows that the United States maintains a consistent domestic collaboration tendency of approximately 0.5, with a gradual increase observed since 2018. In contrast, China's domestic collaboration tendency peaked in 2003 and has since declined, falling below that of the United States after 2010. The tendency for collaboration between the United States and China remains minimal, often approaching negative values, and has further decreased since 2015, indicating a lack of significant collaborative preference between the two countries.
Fig.~\ref{fig:5b} highlights that both China and the United States exhibit a strong preference for citing domestic research. However, China's tendency to cite US research far exceeds the reverse, with the United States showing almost no tendency to cite Chinese research. These patterns of collaboration and citation are consistent across disciplines (Figs.~\ref{fig:5c}--\ref{fig:5d}). 

\subsection{Mobility study}

With the advance of economic globalization and the emergence of a ``big science'' era driven by technological innovation, cross-border mobility has become a defining feature of contemporary scientific labor. We ask whether scholars' international mobility shapes their patterns of collaboration and citation. If so, are these effects primarily driven by the erosion of spatial constraints (a geographic effect) or by the institutional advantages embedded in host countries (a country effect)?

Our previous analysis suggests there is a country-level effect for collaborations and to a lesser extent for citations.
We here consider an alternative method to uncover whether there is a country-level effect.
We make use of the fact that scholars have become increasingly mobile.
This change in country can then be used to study how citations and collaborations are affected by the country.
Mobility is unfortunately unlikely to be exogenous, and is likely to be affected by other factors, including career age, scientific field and the country itself (e.g. through specific mobility programs), that are likely to play a confounding role in collaboration as well.
For this reason, we perform a matching with non-mobile scholars to control for these particular factors.

We considered scholars who moved internationally only once and had at least a career of 5 years before and 5 years after the move.
Following earlier literature on mobility~\cite{robinson-garcia_many_2019}, we label these scholars ``migrants''. In total, we identified 8,179 migrants, with the detailed criteria and the matching approach described in the \nameref{sec:methods}. To capture different transition dynamics, we further categorized them by the number of overlap years between their affiliations in the origin and destination countries, with the distribution shown in Supplementary Table S3.

After filtering, the top five countries of origin for the scholars studied in this article are the US, the UK, Germany, Canada, and Italy. The top five countries of destination are the US, China, the UK, Germany, and Canada (see Supplementary Figure~S4 for distribution).

\begin{figure}[!htbp]
    \centering
    
    \begin{subfigure}[b]{0.48\textwidth}
        \begin{tikzpicture}
            \node[anchor=south west, inner sep=0] (img) at (0,0)
                {\includegraphics[width=\linewidth]{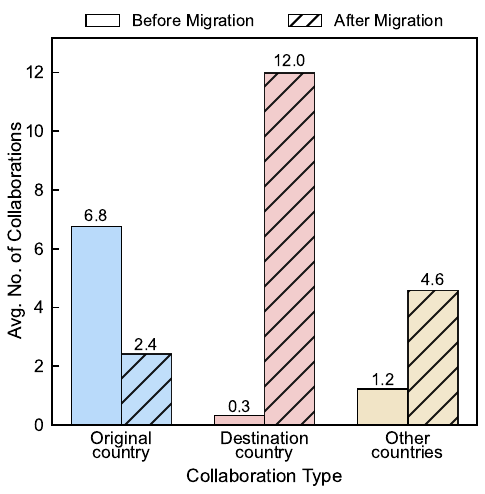}};
            \node[anchor=north west] at (img.north west) {\figlabelsize\textbf{a}};
        \end{tikzpicture}
        \caption{}
        \label{fig:7a}
    \end{subfigure}
    \hfill
    \begin{subfigure}[b]{0.48\textwidth}
        \begin{tikzpicture}
            \node[anchor=south west, inner sep=0] (img) at (0,0)
                {\includegraphics[width=\linewidth]{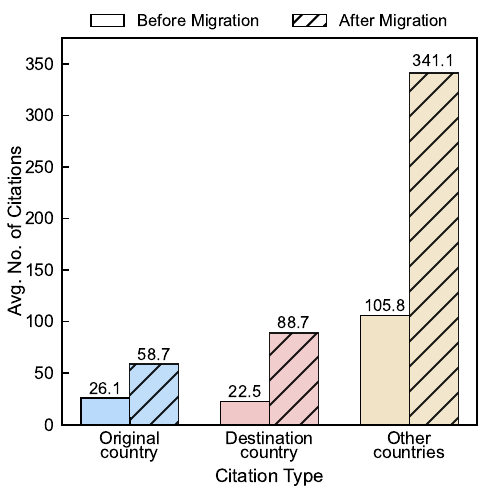}};
            \node[anchor=north west] at (img.north west) {\figlabelsize\textbf{b}};
        \end{tikzpicture}
        \caption{}
        \label{fig:7b}
    \end{subfigure}

    \caption{\textbf{International collaboration and citation patterns of migrant scholars.}
    \textbf{a}, Average number of collaborations with the country of origin, destination, and other countries, before and after migration.
    \textbf{b}, Average number of citations from the country of origin, destination, and other countries, before and after migration. Blank bars indicate collaboration patterns before migration, while hatched bars represent those after migration.}
    \label{fig:mobility}
\end{figure}

We find that the scholars' collaboration patterns undergo a significant shift after mobility (Fig.~\ref{fig:7a}). Before moving, they primarily collaborate with their country of origin, whereas after migration, collaboration with their country of residence becomes predominant. In contrast, the citation patterns exhibit less pronounced change (Fig.~\ref{fig:7b}). The majority of citations come from other countries, with citations from the country of origin being slightly higher than those from the country of residence before migration, while the reverse is true after migration. Results for different overlap years exhibit similar patterns, as shown in Supplementary Figures~S5 and S6.

\begin{figure}[htbp]
  \centering

  \begin{tabular}{@{}c@{\hspace{3mm}}c@{}}
    \begin{subfigure}{0.49\linewidth}
      \centering
      \begin{tikzpicture}
        \node[anchor=south west, inner sep=0] (img) at (0,0) 
          {\includegraphics[width=\linewidth]{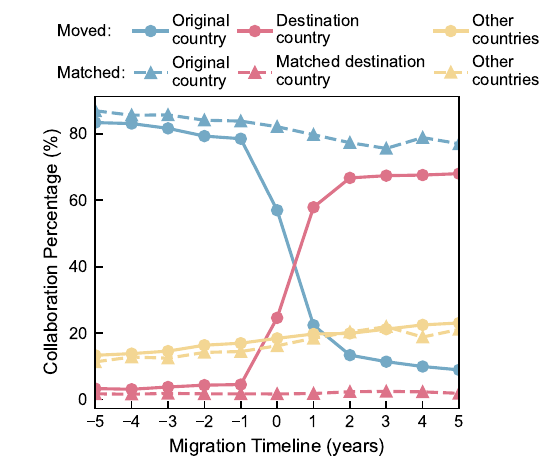}};
        \node[anchor=north west, xshift=5pt] at (img.north west) {\figlabelsize\textbf{a}};
      \end{tikzpicture}
      \caption{}
      \label{fig:match_a}
    \end{subfigure} &
    
    \begin{subfigure}{0.49\linewidth}
      \centering
      \begin{tikzpicture}
        \node[anchor=south west, inner sep=0] (img) at (0,0) 
          {\includegraphics[width=\linewidth]{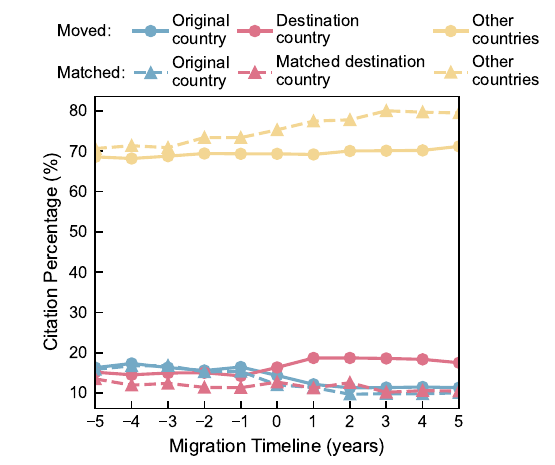}};
        \node[anchor=north west, xshift=5pt] at (img.north west) {\figlabelsize\textbf{b}};
      \end{tikzpicture}
      \caption{}
      \label{fig:match_b}
    \end{subfigure} \\
  \end{tabular}\\[-2.5ex]

  \begin{tabular}{@{}c@{\hspace{3mm}}c@{}}
    
    \begin{subfigure}{0.49\linewidth}
      \centering
      \begin{tikzpicture}
        \node[anchor=south west, inner sep=0] (img) at (0,0) 
          {\includegraphics[width=\linewidth]{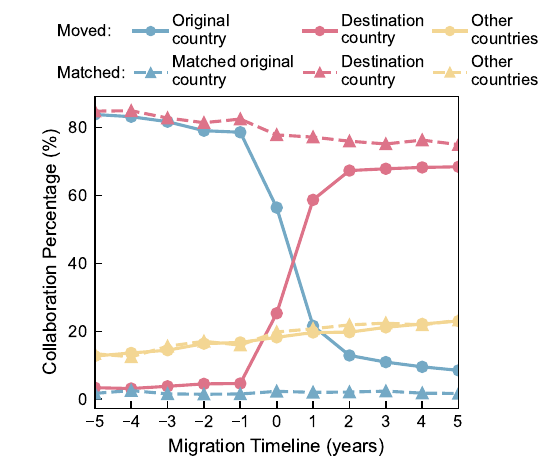}};
        \node[anchor=north west, xshift=5pt] at (img.north west) {\figlabelsize\textbf{c}};
      \end{tikzpicture}
      \caption{}
      \label{fig:match_c}
    \end{subfigure} &

    \begin{subfigure}{0.49\linewidth}
      \centering
      \begin{tikzpicture}
        \node[anchor=south west, inner sep=0] (img) at (0,0) 
          {\includegraphics[width=\linewidth]{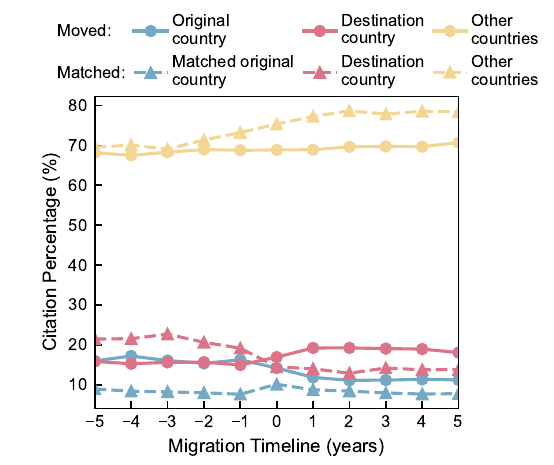}};
        \node[anchor=north west, xshift=5pt] at (img.north west) {\figlabelsize\textbf{d}};
      \end{tikzpicture}
      \caption{}
      \label{fig:match_d}
    \end{subfigure} \\
  \end{tabular}

  \caption{\textbf{Comparison of international collaboration and citation patterns between migrant and non-migrant scholars.}
  \textbf{a}, Collaboration patterns over time for migrant scholars and non-migrant scholars matched on the country of origin. 
  \textbf{b}, Citation patterns over time for migrant scholars and non-migrant scholars matched on the country of origin. 
  \textbf{c}, Collaboration patterns over time for migrant scholars and non-migrant scholars matched on the country of destination. 
  \textbf{d}, Citation patterns over time for migrant scholars and non-migrant scholars matched on the country of destination. Solid lines represent migrant scholars; dashed lines indicate non-migrant scholars matched by country of origin or destination.
}
  \label{fig:mobility_match}
\end{figure}
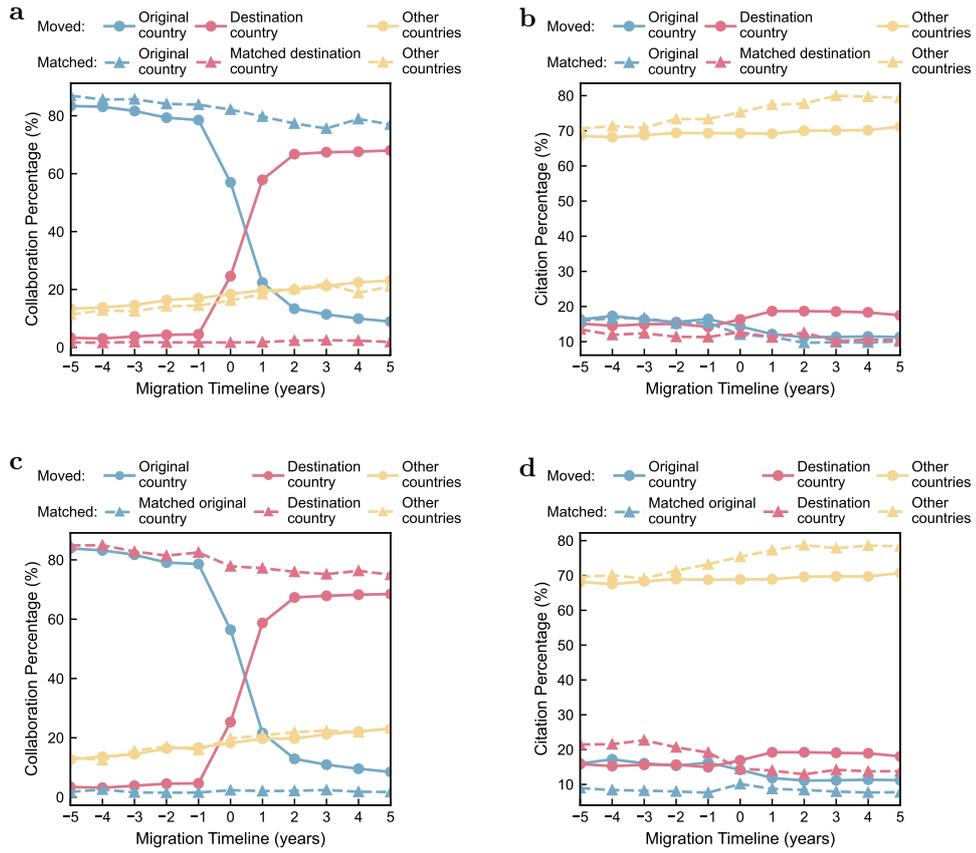

To further examine how these patterns differ from those of non-mobile scholars, we compared migrant scholars with matched non-migrant scholars from both their countries of origin and destination (see section on~\nameref{sec:matching} in the~\nameref{sec:methods}). Compared with non-migrant scholars from the country of origin (Fig.~\ref{fig:match_a}), we observe that before the move year, the collaboration pattern of migrant scholars aligns closely with that of matched non-migrant scholars, where collaboration is primarily with the country of origin, and collaboration with the country of destination is minimal. After the move year, the collaboration pattern of matched non-migrant scholars remains largely unchanged, while the collaboration of migrant scholars with their country of origin sharply declines and quickly stabilizes. In contrast, their collaboration with the destination country gradually increases, maintaining an upward trend. In terms of citations(Fig.~\ref{fig:match_b}), before the move year, both migrant and non-migrant scholars exhibit similar citation patterns, with citations primarily to other countries, and the country of origin being slightly more cited than the country of destination. After the move year, the citation pattern for non-migrant scholars remains unchanged, while migrant scholars slightly increase citations from their country of destination compared to their country of origin. Similar patterns are observed across different overlap years, as shown in Supplementary Figures~S7 and S8.

Next, compared with matched non-migrant scholars from the destination country (Fig.~\ref{fig:match_c}), the results show that before the move year, both migrant and non-migrant scholars primarily collaborate with their respective countries of residence, while non-mobile scholars in the destination country rarely collaborate with the country of origin of the migrant. After the move year, the collaboration pattern of non-migrant scholars in the destination country remains unchanged, while the collaboration pattern of migrant scholars follows the previously observed trend. In terms of citations (Fig.~\ref{fig:match_d}), before the move year, non-migrant scholars from the country of destination predominantly cite other countries, with the country of destination slightly more cited than the migrant scholars’ country of origin. After the move year, the citation pattern for non-migrant scholars remains unchanged, and the citation pattern for migrant scholars is the same as the earlier findings described above. The patterns remain consistent across different overlap years (Supplementary Figures~S9 and S10).

\section{Discussion}

In this study, we examine the impact of geographic distance and country-level factors on international collaboration and citation patterns. Our findings show that the effect of geographic distance on collaboration remains substantial. Our results cast doubt on suggestions that increasing globalization and internationalization of science has gradually diminished the influence of geographic distance on collaboration. In other words, we do not observe any ``death of distance'' \cite{cairncross_death_1997}. Instead, the effect of distance has intensified over time, indicating that geography exerts a persistent influence on collaboration.
In contrast, citation behavior appears to be less sensitive to geography. 
At the country level, we find that collaboration exhibits strong country preferences: most countries tend to collaborate domestically. A similar pattern is observed in citation behavior, although the domestic citations are less pronounced. Notably, many countries tend to cite US research more frequently while under-citing work from China, suggesting the presence of country biases in both collaboration and citation patterns. This imbalance is particularly striking in US–China relations, where citation flows are largely one-directional---from China to the US---but not reciprocated. 

These findings provide critical insights into the mechanisms shaping international collaboration in the digital age. Despite transformative advances in communication technologies, scientific collaboration remains deeply embedded in physical and institutional contexts. Structural factors such as administrative barriers, differences in academic standards, funding systems, cultural differences, and language continue to exert influence that surpasses the convenience of digital connectivity~\cite{hwang2013effects,matthews2020international,vieira2022distance}.

The divergence between collaboration and citation behaviors highlights the distinct logics of knowledge production and knowledge recognition. Collaboration, as a process of co-creating knowledge, is fundamentally shaped by geographic distance, since spatial proximity facilitates frequent face-to-face interaction, fosters trust and social capital, and enables the exchange of tacit knowledge that is difficult to convey across distance~\cite{stephens2021knowledge,van2023impact,chumnangoon2023closeness}. Citation, in contrast, functions as an act of recognition~\cite{merton1973sociology}, that is more globally distributed and shaped more by knowledge flows, intellectual impact, and research visibility than by physical location~\cite{bornmann2008citation,onodera2015factors,tahamtan2016factors}. Evidence from researcher mobility supports this distinction: after relocating, scholars tend to form new collaborations in their host countries rather than maintaining ties with collaborators in their countries of origin. Citation patterns, however, remain relatively stable before and after migration, with most citations continuing to come from international sources.

The observed country preferences, particularly the asymmetry in US–China scientific relations, reflect deeper structural imbalances within the global scientific system. These patterns likely result from multiple interacting mechanisms, including historically entrenched academic path dependencies, the dominance of English in scientific communication, biases in research evaluation systems favoring outputs from certain countries, and geopolitical influences~\cite{gomez2022leading,nielsen2021global,bruck2023bibliometric,nkansah2024dependency}. Together, these factors may lead to the systematic undervaluation of contributions from non-Western countries, reinforcing a ``core–periphery'' structure in global scientific recognition.

Our study has several limitations. 
First, we rely on affiliation data from publication metadata, 
which captures formal, documented collaborations but may miss informal interactions, short-term visits, or other mobility events not reflected in publications. While this means some aspects of scientific collaboration are not observed, using co-authorship data provides a systematic and reliable view of collaboration patterns across a large number of researchers.
Second, we treat all co-authorships and citations equally. This limits our ability to distinguish variation in collaboration depth or citation motivation. Future work could incorporate contribution statements or citation context to refine our analysis. 
Third, although previous studies have shown a strong relationship between international collaboration and citation patterns \cite{pan2012world,wang2024international,duarte2025general}, our study analyzes these two dimensions separately. This approach allows us to disentangle the effects of geographic distance and country-level factors on collaboration and citation, respectively. While this distinction enables clearer attribution of spatial effects, future research could explore the interaction between citations and collaborations more explicitly. 
Fourth, in some cases, our statistical model showed some difficulties when fitting it to data. This problem was particularly visible in the Health Sciences. We currently leave out these results from our analysis. Why the statistical model does not fit well those instances is not clear, and future work could explore alternative models. Moreover, we only modeled citations and collaborations in the top 10\% of all cities. Although this does cover most of the collaborations and citations, future work could explore more efficient estimation procedures to allow including all data.
Finally, although we believe we control for relevant (latent) factors, and present compelling arguments and evidence of country-level effects, it is possible that more elaborate mechanisms underlie these effects. In our current approach, we could interpret these as biases, but it is possible that some mechanisms playing a mediating role. For instance, country-level collaboration may be stimulated explicitly through certain national funding arrangements, so that these effects may perhaps more appropriately be labeled disparities, following~\cite{traag_causal_2022}. Results of our mobility analysis is consistent with such an interpretation: changing countries changes scholars' collaborations, suggesting that these do not constitute personal biases against collaborating with certain nationalities. Future research should clarify this further.

Taken together, our study reveals that geographic distance and country context have a persistent influence on global patterns of scientific collaboration and citation. Our results carry several implications. First, it is important to recognize that digital infrastructure alone cannot ensure an inclusive and collaborative global research ecosystem. While communication technologies facilitate cross-border interactions, collaboration may still be constrained by factors such as institutional differences, administrative barriers, and cultural distance, although these mechanisms were not directly examined in our study. Coordinated policies, including targeted funding, researcher exchanges, and incentives for cross-border partnerships, are therefore essential to promote equitable international collaboration and tackle global scientific challenges.
Second, inequalities in the global scientific system remain pronounced, particularly for scholars from underrepresented regions, whose work often receives lower visibility, fewer citations, and less recognition. This systemic imbalance not only limits the diversity of global knowledge production but also undermines the overall potential of scientific progress. These findings call for the academic community, institutions, and policymakers to critically reflect on existing collaboration and evaluation practices and to promote a more equitable and inclusive global research environment.

\section{Methods}\label{sec:methods}

\subsection{Data}

\subsubsection{Clustering of city locations}
\label{sec:data-cleaning}
One particular problem in OpenAlex is that it includes organizations as belonging to the same city that have quite distinct locations compared to other institutions in the same city.
One very visible example of this are institutions located in St. Denis, which is both a suburb of Paris, but also a town in Île de la Réunion, both of which contain some academic institutions.

We disambiguated these cities by using a clustering approach.
In particular, we considered all organizations within the same city.
We then calculated the geographic distance between these various organizations and clustered the resulting network.
We used the Constant Potts Model~\cite{traag_narrow_2011} optimized by the Leiden algorithm~\cite{traag_louvain_2019} to cluster the network.
We used the distances in km as negative edge weights, while simultaneously using a negative resolution parameter of $\gamma = -30$.
This leads to clusters consisting of locations that are on average located not father than $30$km away from other locations in the same cluster.
We then took the mean positions of all locations in a certain cluster as the average position of the city in the subsequent analysis at the city level.

\subsubsection{Data filtering and inclusion criteria}

We extracted core publications\footnote{https://open.leidenranking.com/information/indicators} from OpenAlex covering the period 2000 to 2022 and restricted our analysis to journal articles. To ensure comparability and analytical robustness, we applied a series of filtering steps. Self-citations were removed from all citation-based indicators. Cities were retained only if they produced at least 50 publications in any given year during the observation window and only countries with at least three qualifying cities were included. 

\subsection{Statistical model}

The challenge with trying to infer a country-level effect for citations and collaborations, is that citations and collaborations are also heavily affected by the underlying propensity to be cited or to collaborate.
These citation and collaboration propensities are not directly visible, but do translate into the observed number of citations and collaborations.
However, if they are also affected by country-level effects at the same time, we cannot simply take those citations and collaborations at face-value.
For that reason, we take a latent variable approach, with the latent variable modeling the citation and collaboration propensity.

In general, we model citations and collaborations as a Poisson distribution
\begin{equation}
  w_{ij} \sim \text{Poisson}(\lambda_{ij}),  
\end{equation}
with $\lambda_{ij}$ the overall rate of citation and collaboration.
We define $\lambda_{ij}$ for collaboration as

\begin{equation}
    \lambda_{ij} = \alpha \frac{n_i \theta_i n_j \theta_j}{d_{ij}^\beta} C_{c_i,c_j},    
\end{equation}
while $\lambda_{ij}$ for citations is defined as

\begin{equation}
    \lambda_{ij}(t) = \alpha f(t)\frac{n_i n_j \theta_j}{d_{ij}^\beta} C_{c_i,c_j},
\end{equation}
where $f(t)$ is the lognormal density for $t = y_{\text{citing}} - y_{\text{cited}}$, following \cite{yin_time_2017}.
That is, for some $\mu$ and $\sigma$, $f(t)$ equals
\begin{equation}
    f(t) = \frac {1}{ t\sigma \sqrt{2\pi} }\exp \left(-{\frac {\left(\ln t-\mu \ \right)^{2}}{2\sigma ^{2}}}\right).
\end{equation}

Without further constraints, the specified parameters are not identifiable, due to degeneracies between $\theta_i$, $C$ and $\alpha$.
We therefore impose further constraints such that $\prod_i \theta_i = 1$ and $\prod_s C_{rs} = 1$ for each $r$.
We constrain the geometric average, because that is more convenient in the implementation at a logarithmic scale, because this translates into a sum-to-zero constraints (i.e. $\sum_i \log \theta_i = 0$).
The overall rate of citation/collaboration is then estimated by $\alpha$, while city-level differences are estimated by $\theta_i$, while specific country-level effect should be properly estimated by $C$.

\subsection{Bayesian approach}

We have modeled this in a Bayesian model, with informative priors on the various parameters.
We assume $\beta \geq 0$ and use a half-normal prior $\beta \sim \text{Normal}(0, 1)$.
We assume a relatively broad prior on $\log \alpha \sim \text{Normal}(-10, 20)$.
This is centered on the negative side, since the number of collaborations is typically considerably less than the full combination of number of publications $n_i n_j$.
Finally, we consider more narrow priors on collaboration of $\log C_{rs} = \text{Normal}(0, 1/n_{\text{countries}})$ and $\log \theta_i = \text{Normal}(0, 1/n_{\text{cities}})$.
Hence, there needs to be relatively strong evidence for these estimates to be swayed from their prior around $0$.

\subsection{Identification of migrant scholars}
Based on existing classifications of mobility types \cite{robinson-garcia_many_2019}, we define migrants as scholars who, during an initial period $t_0$ to $t_a$, were affiliated exclusively with Country A (based on their institutional affiliations listed in publications), and who, in a subsequent period $t_b$, were affiliated exclusively with Country B. Scholars with affiliations to any third country between these two periods are excluded from the analysis to ensure a clean country-to-country migration trajectory.

To ensure that only established and consistently publishing scholars were included in the migration analysis, we applied additional inclusion criteria. We restricted the sample to scholars whose first publication occurred in or after 2000. Each scholar was required to have a publication record of at least five years before migration, with a minimum of three publications during that period. Similarly, each scholar was required to have a publication record of at least five years after migration, also with a minimum of three publications in that post-migration phase. These filters ensure sufficient data for analyzing changes in collaboration and citation patterns before and after migration.

To analyze the temporal dynamics of migration, we introduce the concept of the move year $t_m$, which represents the estimated year a scholar migrated from Country A to Country B. The move year is determined by the publication history and affiliation patterns, specifically the last year of affiliation with Country A (
$t_a$) and the first year of affiliation with Country B ($t_b$).
We define two core migration scenarios based on the relationship between $t_a$ and $t_b$:

\begin{itemize}
    \item \textbf{No Overlap Years:} If $t_a + 1 \leq t_b$, the migrant exhibits no overlapping affiliations between the two countries. This scenario encompasses all cases from direct moves (no gap) to those with a gap of up to two years.

    \item \textbf{Overlap Years:} If $t_a + 1 > t_b$, the migrant has overlapping affiliations in both countries. We restrict such overlaps to a maximum of two years. 
\end{itemize}
For all scenarios, we estimate the move year $t_m$ using a unified formula that calculates the midpoint between $t_a$ and $t_b$:
\begin{equation}
    t_m = \left\lfloor t_a + \frac{|t_b - t_a|}{2} \right\rfloor
\end{equation}
Based on the number of overlap years, we categorize migrants into three groups: 0, 1, or 2 overlap years. Scholars with more than two years of overlap are excluded to maintain consistency and data reliability.

\subsection{Matching methodology}
\label{sec:matching}
Let $\mathcal{M}$ denote the set of all ``migrant'' researchers who relocated internationally at least once.  For each migrant $i \in \mathcal{M}$, let $Y_i$ be the year of their first publication, $O_i$ their original country (i.e. the affiliation in year $Y_i$), and $D_i$ their destination country (i.e. the affiliation in the first year after relocation).  Let $\mathcal{C}_i$ be the pool of non‑migrant researchers whose first publication year satisfies $Y_j = Y_i$.

For each $i \in \mathcal{M}$, we establish two independent strata of eligible one‑to‑one matches with non-migrant researchers $j \in \mathcal{C}_i$:

\begin{enumerate}
  \item \textbf{Original‑country match.}  Choose a control $j \in \mathcal{C}_i$ such that $ O_{j} = O_i $.  
  \item \textbf{Destination‑country match.}  Choose a control $j \in \mathcal{C}_i$ such that $ D_{j} = D_i$.
\end{enumerate}

Within each stratum (original or destination), we rank eligible controls by a composite distance $\Delta(j,i)$ that combines pre‑relocation publication volume and research topic similarity.
We first consider the similarity in cumulative publication volume up to the move year. For a migrant scholar $i$ with move year $t_m$, we compare their publication count up to $t_m$ with that of a control scholar $j$ up to the same calendar year $t_m$.
In particular, let $P_k(t)$ be the set of publications of researcher $k$ up to, but not including, year $t$.
We then define the difference in overall publication volume as 
\begin{equation}
    \delta_p(j,i) = \left|\; | P_j(t_m) | - | P_i(t_m) | \; \right|.
\end{equation}

Research topic similarity is based on OpenAlex’s \texttt{concept} field with weights. To capture a scholar's research profile at a comparable career stage, we first construct a concept vector for each researcher $k$. This vector, denoted as $v_k(c)$, aggregates their exposure to concept $c$ prior to a baseline year. For each researcher $k$, we aggregate concept weights before the move year $t_m$:
\begin{equation}
  v_k(c) = \sum_{p \in P_k(t_m)} w_p(c),
\end{equation}
where $w_p(c)$ is the weight of concept $c$ in paper $p$.  We then define
\begin{equation}
  S_c(j,i) = \sum_{c} v_j(c)\,v_i(c),
\end{equation}
so that only shared concepts contribute (each contributes $w_j(c)\times w_i(c)$).

Finally, we normalize and combine the publication volume similarity and the research topic similarity:
\begin{equation}
  \Delta(j,i) =
    \beta_p \frac{\delta_p(j,i)}{\max_{j'}\delta_p(j',i)}
    +
    \beta_c \Bigl(1 - \frac{S_c(j,i)}{\max_{j'}S_c(j',i)}\Bigr),
\end{equation}
with $\beta_p + \beta_c = 1$.
The weights $\beta_p$ and $\beta_c$ can be adjusted depending on the desired emphasis on productivity versus topical similarity. In our analysis, we set $\beta_p = \beta_c = 0.5$, giving equal weight to both criteria.

For each $i$, we select the control $j$ that minimizes $\Delta(j,i)$, yielding one non‑migrant peer matched on original country ($O_i$) and one matched on destination country ($D_i$).  This procedure ensures exact alignment on first publication year and country while maximizing similarity in pre‑relocation output and research focus, thereby controlling for career length, country, productivity and field.

\section*{Data availability}

All data used in this study are available in a public repository. 
The replication materials, include datasets used for the modelling analysis and for the mobility matching analysis, are deposited at Zenodo and can be accessed via the following DOI: \url{https://doi.org/10.5281/zenodo.18018927}. 

\bibliography{mybib}

@article{yin_time_2017,
	title = {The {Time} {Dimension} of {Science}: {Connecting} the {Past} to the {Future}},
	volume = {11},
	issn = {17511577},
	shorttitle = {The {Time} {Dimension} of {Science}},
	url = {https://linkinghub.elsevier.com/retrieve/pii/S1751157717300020},
	doi = {10.1016/j.joi.2017.04.002},
	language = {en},
	number = {2},
	urldate = {2023-03-14},
	journal = {Journal of Informetrics},
	author = {Yin, Yian and Wang, Dashun},
	month = may,
	year = {2017},
	note = {arXiv:1704.04657 [physics]},
	pages = {608--621},
}

@article{bornmann2008citation,
  title={What do citation counts measure? A review of studies on citing behavior},
  author={Bornmann, Lutz and Daniel, Hans-Dieter},
  journal={Journal of documentation},
  volume={64},
  number={1},
  pages={45--80},
  year={2008},
  publisher={Emerald Group Publishing Limited}
}

@article{tahamtan2016factors,
  title={Factors affecting number of citations: a comprehensive review of the literature},
  author={Tahamtan, Iman and Safipour Afshar, Askar and Ahamdzadeh, Khadijeh},
  journal={Scientometrics},
  volume={107},
  pages={1195--1225},
  year={2016},
  publisher={Springer}
}

@article{newman2004coauthorship,
  title={Coauthorship networks and patterns of scientific collaboration},
  author={Newman, Mark EJ},
  journal={Proceedings of the national academy of sciences},
  volume={101},
  number={suppl\_1},
  pages={5200--5205},
  year={2004},
  publisher={National Academy of Sciences}
}

@article{capello2018proximities,
  title={Proximities and the intensity of scientific relations: synergies and nonlinearities},
  author={Capello, Roberta and Caragliu, Andrea},
  journal={International Regional Science Review},
  volume={41},
  number={1},
  pages={7--44},
  year={2018},
  publisher={SAGE Publications Sage CA: Los Angeles, CA}
}

@article{yan2012scholarly,
  title={Scholarly network similarities: How bibliographic coupling networks, citation networks, cocitation networks, topical networks, coauthorship networks, and coword networks relate to each other},
  author={Yan, Erjia and Ding, Ying},
  journal={Journal of the American Society for Information Science and Technology},
  volume={63},
  number={7},
  pages={1313--1326},
  year={2012},
  publisher={Wiley Online Library}
}

@article{matthews2020international,
  title={International scientific collaborative activities and barriers to them in eight societies},
  author={Matthews, Kirstin RW and Yang, Erin and Lewis, Steven W and Vaidyanathan, Brandon R and Gorman, Monica},
  journal={Accountability in Research},
  volume={27},
  number={8},
  pages={477--495},
  year={2020},
  publisher={Taylor \& Francis}
}

@article{wang2024international,
  title={International collaboration leading to high citations: Global impact or home country effect?},
  author={Wang, Jue and Frietsch, Rainer and Neuh{\"a}usler, Peter and Hooi, Rosalie},
  journal={Journal of Informetrics},
  volume={18},
  number={4},
  pages={101565},
  year={2024},
  publisher={Elsevier}
}

@article{robinson-garcia_many_2019,
	title = {The many faces of mobility: {Using} bibliometric data to measure the movement of scientists},
	volume = {13},
	issn = {1751-1577},
	shorttitle = {The many faces of mobility},
	doi = {10.1016/j.joi.2018.11.002},
	number = {1},
	urldate = {2024-05-16},
	journal = {Journal of Informetrics},
	author = {Robinson-Garcia, Nicolás and Sugimoto, Cassidy R. and Murray, Dakota and Yegros-Yegros, Alfredo and Larivière, Vincent and Costas, Rodrigo},
	month = feb,
	year = {2019},
	pages = {50--63},
}

@article{isfandyari2023global,
  title={Global scientific collaboration: A social network analysis and data mining of the co-authorship networks},
  author={Isfandyari-Moghaddam, Alireza and Saberi, Mohammad Karim and Tahmasebi-Limoni, Safieh and Mohammadian, Sajjad and Naderbeigi, Farahnaz},
  journal={Journal of Information Science},
  volume={49},
  number={4},
  pages={1126--1141},
  year={2023},
  publisher={SAGE Publications Sage UK: London, England}
}

@article{gomez2022leading,
  title={Leading countries in global science increasingly receive more citations than other countries doing similar research},
  author={Gomez, Charles J and Herman, Andrew C and Parigi, Paolo},
  journal={Nature Human Behaviour},
  volume={6},
  number={7},
  pages={919--929},
  year={2022},
  publisher={Nature Publishing Group UK London}
}

@article{lariviere2018citations,
  title={Citations strength begins at home},
  author={Lariviere, Vincent and Gong, Kaile and Sugimoto, Cassidy R},
  journal={Nature},
  volume={564},
  number={7735},
  pages={S70--S70},
  year={2018},
  publisher={Nature Publishing Group}
}

@article{nielsen2021global,
  title={Global citation inequality is on the rise},
  author={Nielsen, Mathias Wullum and Andersen, Jens Peter},
  journal={Proceedings of the National Academy of Sciences},
  volume={118},
  number={7},
  pages={e2012208118},
  year={2021},
  publisher={National Academy of Sciences}
}

@article{pan2012world,
  title={World citation and collaboration networks: uncovering the role of geography in science},
  author={Pan, Raj Kumar and Kaski, Kimmo and Fortunato, Santo},
  journal={Scientific reports},
  volume={2},
  number={1},
  pages={902},
  year={2012},
  publisher={Nature Publishing Group UK London}
}

@article{bassecoulard2000insights,
  title={Insights in determinants of international scientific cooperation},
  author={Bassecoulard, ELISE and Okubo, YOSHIKO and Zitt, MICHEL},
  journal={Collaboration in Science and in Technology},
  pages={13},
  year={2000}
}

@article{okubo2004searching,
  title={Searching for research integration across Europe: a closer look at international and inter-regional collaboration in France},
  author={Okubo, Yoshiko and Zitt, Michel},
  journal={Science and Public Policy},
  volume={31},
  number={3},
  pages={213--226},
  year={2004},
  publisher={Beech Tree Publishing}
}

@article{leonard2012cooperation,
  title={Cooperation across cultures: An examination of the concept in 16 countries},
  author={Leonard, Karen Moustafa and Cosans, Christopher and Pakdil, Fatma and Collaborator, Country},
  journal={International Journal of Intercultural Relations},
  volume={36},
  number={2},
  pages={238--247},
  year={2012},
  publisher={Elsevier}
}

@misc{traag_causal_2022,
	title = {Causal foundations of bias, disparity and fairness},
	url = {http://arxiv.org/abs/2207.13665},
	doi = {10.48550/arXiv.2207.13665},
	urldate = {2023-04-10},
	publisher = {arXiv},
	author = {Traag, V. A. and Waltman, L.},
	month = dec,
	year = {2022},
}

@book{cairncross_death_1997,
	address = {Boston, Mass},
	title = {The {Death} of {Distance}: {How} the {Communications} {Revolution} {Will} {Change} {Our} {Lives}},
	isbn = {978-0-87584-806-8},
	shorttitle = {The {Death} of {Distance}},
	language = {English},
	publisher = {Harvard Business Review Press},
	author = {Cairncross, Frances},
	year = {1997},
}

@article{traag_narrow_2011,
	title = {Narrow scope for resolution-limit-free community detection},
	volume = {84},
	copyright = {All rights reserved},
	issn = {1539-3755},
	url = {http://arxiv.org/abs/1104.3083},
	doi = {10.1103/PhysRevE.84.016114},
	number = {1},
	journal = {Physical Review E},
	author = {Traag, V. A. and Van Dooren, P. and Nesterov, Y.},
	month = jul,
	year = {2011},
	pages = {016114}
}

@article{traag_louvain_2019,
	title = {From {Louvain} to {Leiden}: guaranteeing well-connected communities},
	volume = {9},
	copyright = {2019 The Author(s)},
	issn = {2045-2322},
	shorttitle = {From {Louvain} to {Leiden}},
	url = {https://www.nature.com/articles/s41598-019-41695-z},
	doi = {10.1038/s41598-019-41695-z},
	language = {en},
	number = {1},
	urldate = {2022-11-02},
	journal = {Scientific Reports},
	author = {Traag, V. A. and Waltman, L. and van Eck, N. J.},
	month = mar,
	year = {2019},
	note = {Number: 1
Publisher: Nature Publishing Group},
	pages = {5233}
}

@article{chumnangoon2023closeness,
  title={How closeness matters: The role of geographical proximity in social capital development and knowledge sharing in SMEs},
  author={Chumnangoon, Pollawat and Chiralaksanakul, Anukal and Chintakananda, Asda},
  journal={Competitiveness Review: An International Business Journal},
  volume={33},
  number={2},
  pages={280--301},
  year={2023},
  publisher={Emerald Publishing Limited}
}

@book{merton1973sociology,
  title={The sociology of science: Theoretical and empirical investigations},
  author={Merton, Robert K},
  year={1973},
  publisher={University of Chicago press}
}

@article{onodera2015factors,
  title={Factors affecting citation rates of research articles},
  author={Onodera, Natsuo and Yoshikane, Fuyuki},
  journal={Journal of the Association for Information Science and Technology},
  volume={66},
  number={4},
  pages={739--764},
  year={2015},
  publisher={Wiley Online Library}
}

@article{stephens2021knowledge,
  title={Knowledge creation through collaboration: The role of shared institutional affiliations and physical proximity},
  author={Stephens, Bryan and Cummings, Jonathon N},
  journal={Journal of the Association for Information Science and Technology},
  volume={72},
  number={11},
  pages={1337--1353},
  year={2021},
  publisher={Wiley Online Library}
}

@article{bruck2023bibliometric,
  title={A bibliometric analysis of geographic disparities in the authorship of leading medical journals},
  author={Br{\"u}ck, Oscar},
  journal={Communications Medicine},
  volume={3},
  number={1},
  pages={178},
  year={2023},
  publisher={Nature Publishing Group UK London}
}

@article{nkansah2024dependency,
  title={Dependency and neocolonialism in international research collaboration: evidence from a Ghanaian elite university},
  author={Nkansah, Jacob Oppong and Oldac, Yusuf Ikbal and Yang, Lili},
  journal={Higher Education},
  pages={1--19},
  year={2024},
  publisher={Springer}
}

@article{hwang2013effects,
  title={Effects of the language barrier on processes and performance of international scientific collaboration, collaborators’ participation, organizational integrity, and interorganizational relationships},
  author={Hwang, Kumju},
  journal={Science Communication},
  volume={35},
  number={1},
  pages={3--31},
  year={2013},
  publisher={SAGE Publications Sage CA: Los Angeles, CA}
}

@article{vieira2022distance,
  title={Which distance dimensions matter in international research collaboration? A cross-country analysis by scientific domain},
  author={Vieira, Elizabeth S and Cerdeira, Jorge and Teixeira, Aurora AC},
  journal={Journal of Informetrics},
  volume={16},
  number={2},
  pages={101259},
  year={2022},
  publisher={Elsevier}
}

@article{frenken2020geography,
  title={Geography of scientific knowledge: A proximity approach},
  author={Frenken, Koen},
  journal={Quantitative Science Studies},
  volume={1},
  number={3},
  pages={1007--1016},
  year={2020},
  publisher={MIT Press One Rogers Street, Cambridge, MA 02142-1209, USA journals-info~…}
}

@article{wang2021knowledge,
  title={Knowledge flows from public science to industrial technologies},
  author={Wang, Lili and Li, Zexia},
  journal={The Journal of technology transfer},
  volume={46},
  number={4},
  pages={1232--1255},
  year={2021},
  publisher={Springer}
}

@article{duarte2025general,
  title={A general analysis of the impact of international collaboration on the citation indices of scientific publications from 60 institutions across five continents},
  author={Duarte, Iraine and Santos, Hellen GG and Rodrigues, Marcio L},
  journal={Anais da Academia Brasileira de Ci{\^e}ncias},
  volume={97},
  number={1},
  pages={e20241035},
  year={2025},
  publisher={SciELO Brasil}
}

@article{megnigbeto2015effect,
  title={Effect of international collaboration on knowledge flow within an innovation system: a Triple Helix approach},
  author={M{\^e}gnigb{\^e}to, Eustache},
  journal={Triple Helix},
  volume={2},
  number={1},
  pages={1--21},
  year={2015},
  publisher={Brill}
}

@article{zhuge2006discovery,
  title={Discovery of knowledge flow in science},
  author={Zhuge, Hai},
  journal={Communications of the ACM},
  volume={49},
  number={5},
  pages={101--107},
  year={2006},
  publisher={ACM New York, NY, USA}
}

@article{nelson2009measuring,
  title={Measuring knowledge spillovers: What patents, licenses and publications reveal about innovation diffusion},
  author={Nelson, Andrew J},
  journal={Research policy},
  volume={38},
  number={6},
  pages={994--1005},
  year={2009},
  publisher={Elsevier}
}

@article{singh2005collaborative,
  title={Collaborative networks as determinants of knowledge diffusion patterns},
  author={Singh, Jasjit},
  journal={Management science},
  volume={51},
  number={5},
  pages={756--770},
  year={2005},
  publisher={INFORMS}
}

@article{jaffe1993geographic,
  title={Geographic localization of knowledge spillovers as evidenced by patent citations},
  author={Jaffe, Adam B and Trajtenberg, Manuel and Henderson, Rebecca},
  journal={the Quarterly journal of Economics},
  volume={108},
  number={3},
  pages={577--598},
  year={1993},
  publisher={MIT Press}
}

@article{sorenson2007science,
  title={Science, social networks and spillovers},
  author={Sorenson, Olav and Singh, Jasjit},
  journal={Industry and Innovation},
  volume={14},
  number={2},
  pages={219--238},
  year={2007},
  publisher={Taylor \& Francis}
}

@article{sidone2017scholarly,
  title={Scholarly publication and collaboration in B razil: The role of geography},
  author={Sidone, Ot{\'a}vio Jos{\'e} Guerci and Haddad, Eduardo Amaral and Mena-Chalco, Jes{\'u}s Pascual},
  journal={Journal of the Association for Information Science and Technology},
  volume={68},
  number={1},
  pages={243--258},
  year={2017},
  publisher={Wiley Online Library}
}

@article{toobaee2024proximity,
  title={Proximity Matters: Analyzing the Role of Geographical Proximity in Shaping AI Research Collaborations},
  author={Toobaee, Mohammadmahdi and Schiffauerova, Andrea and Ebadi, Ashkan},
  journal={arXiv preprint arXiv:2406.06662},
  year={2024}
}

@article{van2023impact,
  title={The impact of geographical distance on learning through collaboration},
  author={Van der Wouden, Frank and Youn, Hyejin},
  journal={Research Policy},
  volume={52},
  number={2},
  pages={104698},
  year={2023},
  publisher={Elsevier}
}

@article{miao2024persistent,
  title={Persistent hierarchy in contemporary international collaboration},
  author={Miao, Lili and Larivi{\`e}re, Vincent and Lee, Byungkyu and Ahn, Yong-Yeol and Sugimoto, Cassidy R},
  journal={arXiv preprint arXiv:2410.13020},
  year={2024}
}

@article{nettasinghe2021emergence,
  title={Emergence of structural inequalities in scientific citation networks},
  author={Nettasinghe, Buddhika and Alipourfard, Nazanin and Krishnamurthy, Vikram and Lerman, Kristina},
  journal={arXiv preprint arXiv:2103.10944},
  year={2021}
}

@incollection{frenken2019death,
  title={Death of distance in science? A gravity approach to research collaboration},
  author={Frenken, Koen and Hoekman, Jarno and Kok, Suzanne and Ponds, Roderik and van Oort, Frank and van Vliet, Joep},
  booktitle={Innovation networks: New approaches in modelling and analyzing},
  pages={43--57},
  year={2019},
  publisher={Springer}
}

@article{paris1998region,
  title={Region-based citation bias in science},
  author={Paris, Gianmarco and De Leo, Giulio and Menozzi, Paolo and Gatto, Marino},
  journal={Nature},
  volume={396},
  number={6708},
  pages={210--210},
  year={1998},
  publisher={Nature Publishing Group UK London}
}

@article{meneghini2008articles,
  title={Articles by Latin American authors in prestigious journals have fewer citations},
  author={Meneghini, Rogerio and Packer, Abel L and Nassi-Calo, Lilian},
  journal={Plos one},
  volume={3},
  number={11},
  pages={e3804},
  year={2008},
  publisher={Public Library of Science San Francisco, USA}
}

@article{akre2011differences,
  title={Differences in citation rates by country of origin for papers published in top-ranked medical journals: do they reflect inequalities in access to publication?},
  author={Akre, Olof and Barone-Adesi, Francesco and Pettersson, Andreas and Pearce, Neil and Merletti, Franco and Richiardi, Lorenzo},
  journal={Journal of Epidemiology \& Community Health},
  volume={65},
  number={2},
  pages={119--123},
  year={2011},
  publisher={BMJ Publishing Group Ltd}
}

\appendix

\section{Supplementary Information}
\label{sec:online_appendix}

\pagenumbering{arabic}
\renewcommand*{\thepage}{S\arabic{page}}
\setcounter{page}{1}

\noindent
\textbf{Content} \hfill \textbf{Page}\\[0.5em]

\noindent\textbf{Figures}\\[0.3em]
Figure S.1. Collaboration preference between countries by year \dotfill S2\\
Figure S.2. Citation preference between countries by year \dotfill S4\\
Figure S.3. Collaboration and citation preferences across research fields \dotfill S6\\
Figure S.4. Origin and destination countries of migrant scholars\dotfill S7\\ 
Figure S.5. Comparison of international collaboration patterns by overlap (0, 1, 2) \dotfill S8\\
Figure S.6. Comparison of international citation patterns by overlap (0, 1, 2) \dotfill S8\\
Figure S.7. Comparison of collaboration patterns over time by overlap (original side) \dotfill S9\\
Figure S.8. Comparison of collaboration patterns over time by overlap (destination side) \dotfill S9\\
Figure S.9. Comparison of citation patterns over time by overlap (original side) \dotfill S10\\
Figure S.10. Comparison of citation patterns over time by overlap (destination side) \dotfill S10\\[1em]

\noindent\textbf{Tables}\\[0.3em]
Table S.1. Mapping of openAlex concepts to field names \dotfill S11\\
Table S.2. Mapping of country codes to country names \dotfill S11\\
Table S.3. Number of scholars by migration scenario \dotfill S12\\[1em]

\clearpage


\begin{figure}[p]
  \centering
  \includegraphics[width=\textwidth]{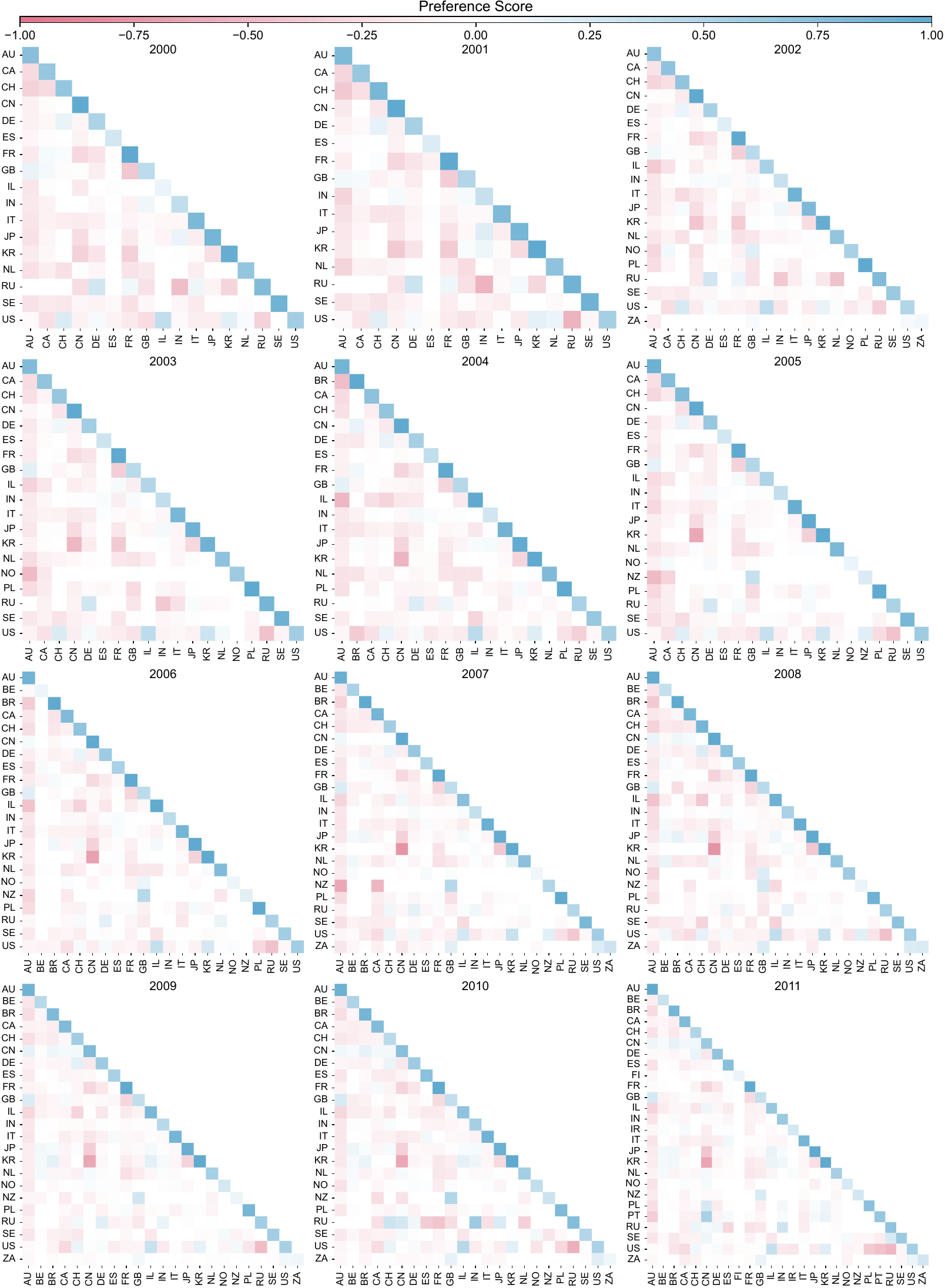}
  \caption*{\textbf{\textit{Figure S.1(a).} Collaboration preference between countries by year.} Overall collaboration patterns across all disciplines are shown.
Color intensity indicates collaboration strength — lighter pink denotes lower preference, deeper blue indicates higher preference.}
  \label{fig:s1a}
\end{figure}

\begin{figure}[p]
  \centering
  \includegraphics[width=\textwidth]{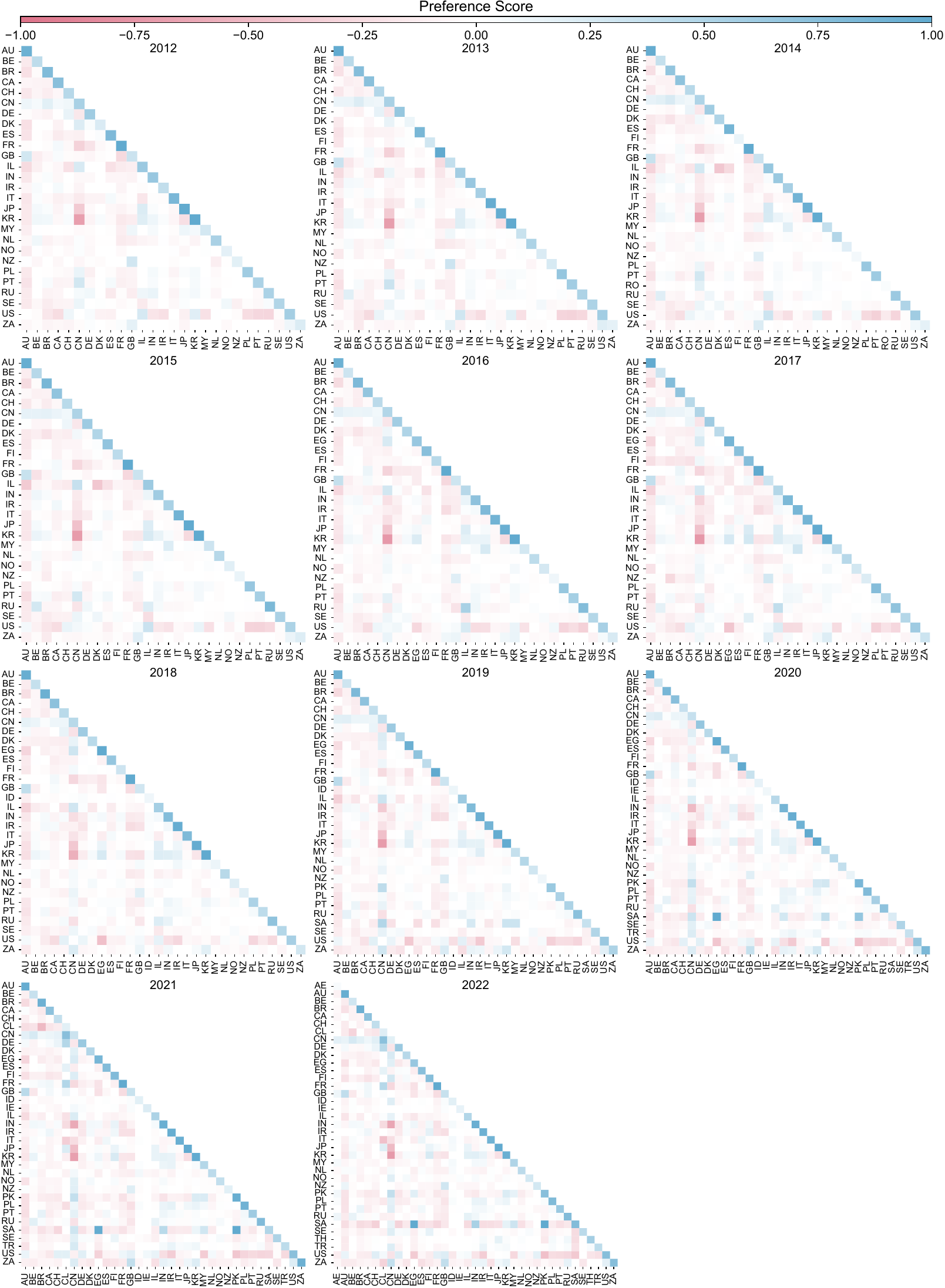}
  \caption*{\textbf{\textit{Figure S.1(b, continued).} Collaboration preference between countries by year.} Overall collaboration patterns across all disciplines are shown (continued).}
  \label{fig:s1b}
\end{figure}

\begin{figure}[p]
  \centering
  \includegraphics[width=\textwidth]{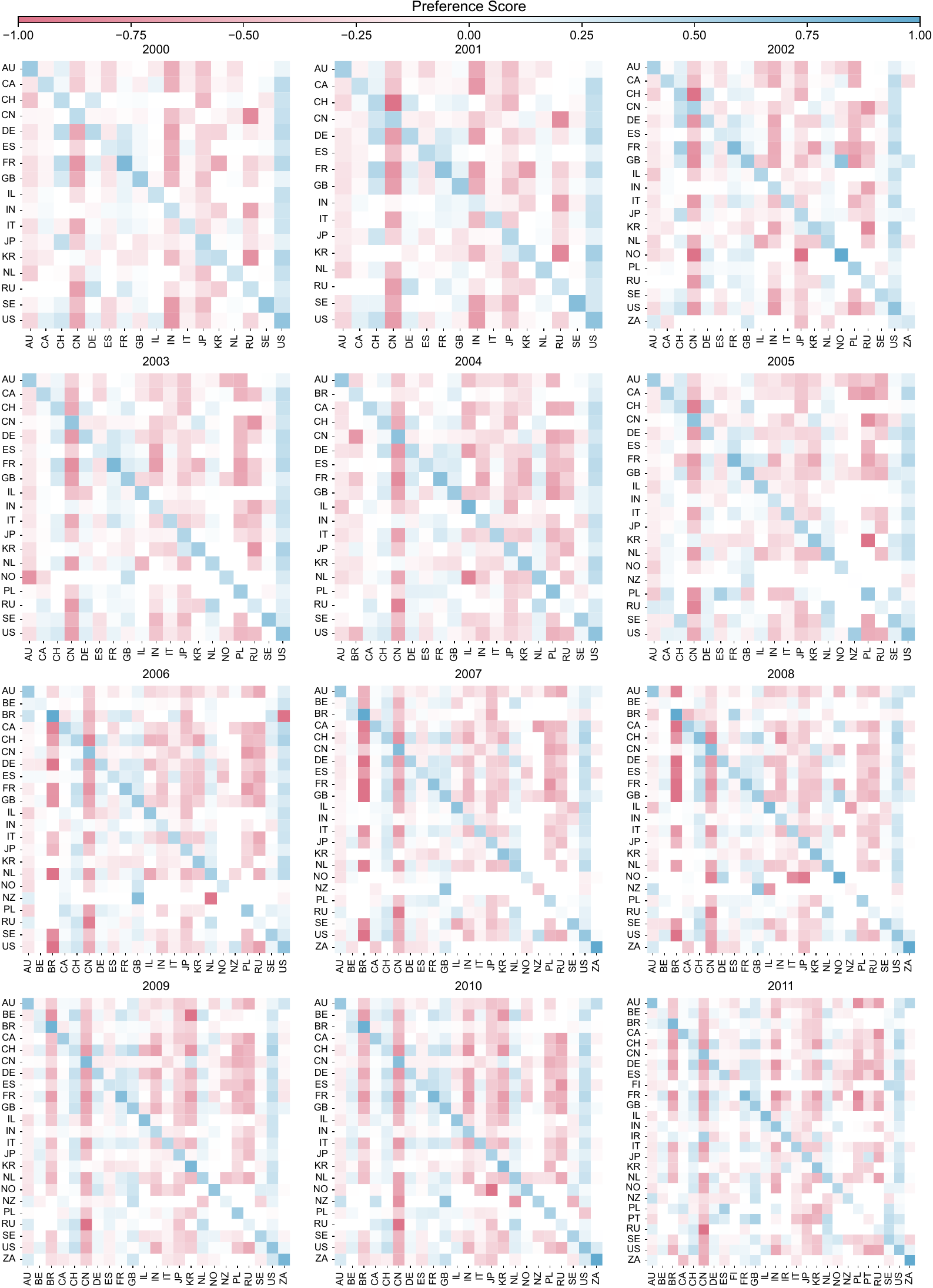}
  \caption*{\textbf{\textit{Figure S.2(a).} Citation preference between countries by year.} Overall citation patterns across all disciplines are shown.
Color intensity indicates citation strength — lighter pink denotes lower preference, deeper blue indicates higher preference.}
  \label{fig:s2a}
\end{figure}

\begin{figure}[p]
  \centering
  \includegraphics[width=\textwidth]{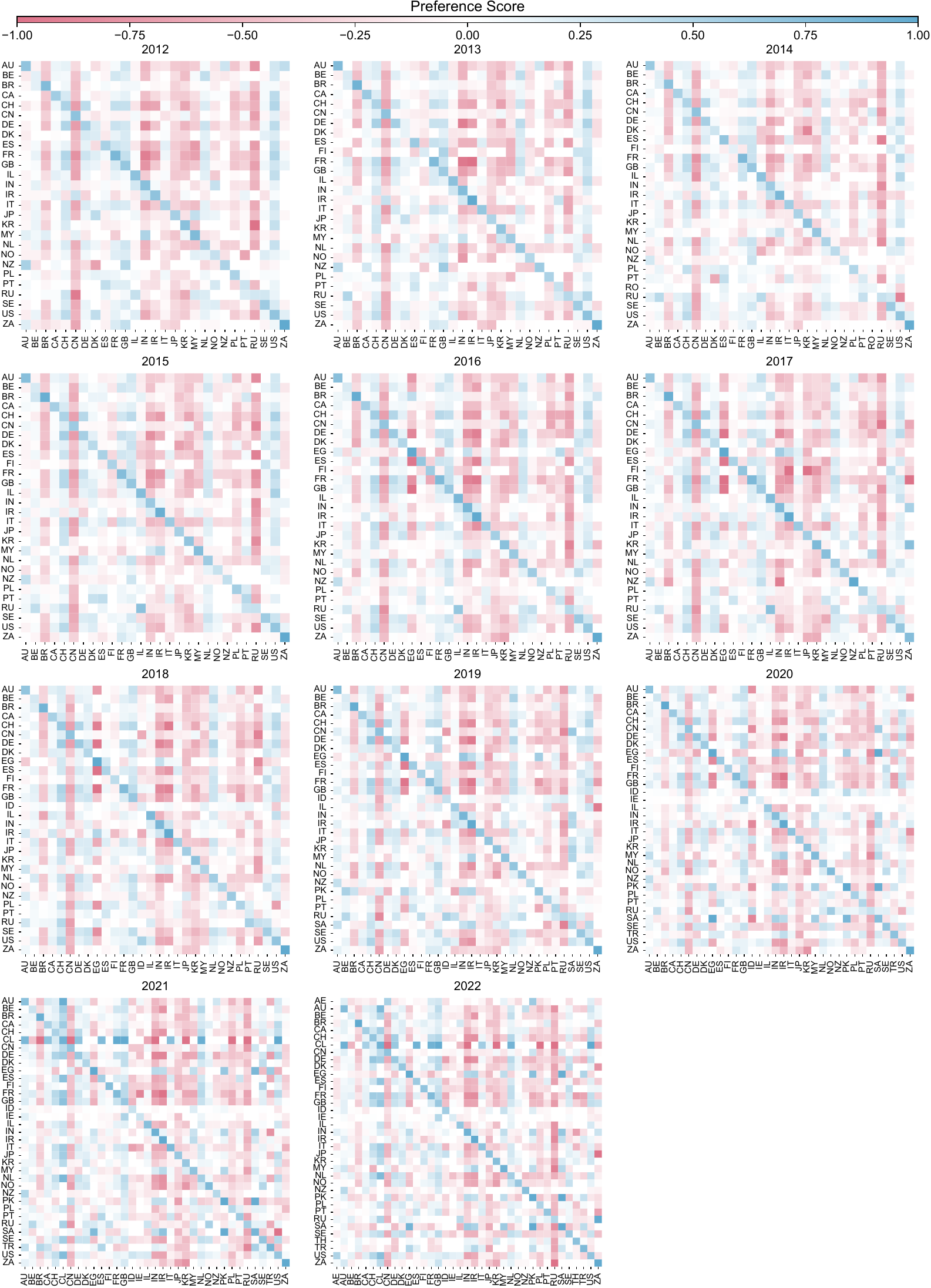}
  \caption*{\textbf{\textit{Figure S.2(b, continued).} Citation preference between countries by year.} Overall citation patterns across all disciplines are shown (continued).}
  \label{fig:s2b}
\end{figure}

\begin{figure}[!htbp]
    \centering

    \begin{subfigure}[a]{\textwidth}
        \centering
        \begin{tikzpicture}
            \node[inner sep=0] (img)
                {\includegraphics[width=\linewidth]{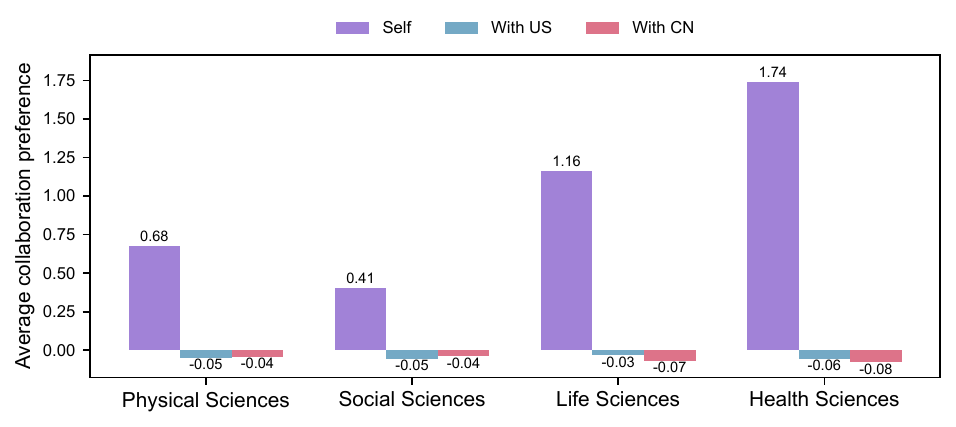}};
            \node[anchor=north west, xshift=0pt, yshift=0pt] at (img.north west) {\figlabelsize\textbf{a}};
        \end{tikzpicture}
        \label{fig:combined_sa}
    \end{subfigure}

    \vspace{-14pt}

    \begin{subfigure}[b]{\textwidth}
        \centering
        \begin{tikzpicture}
            \node[inner sep=0] (img)
                {\includegraphics[width=\linewidth]{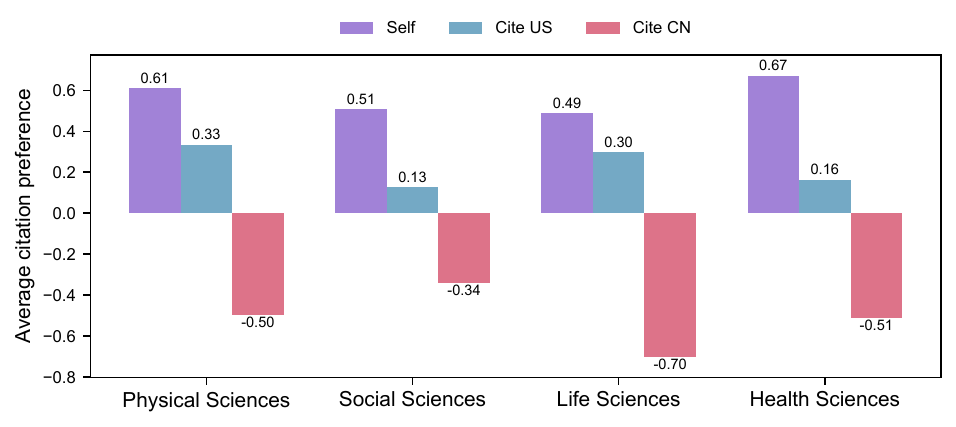}};
            \node[anchor=north west, xshift=0pt, yshift=0pt] at (img.north west) {\figlabelsize\textbf{b}};
        \end{tikzpicture}
        \label{fig:combined_sb}
    \end{subfigure}
    
    \vspace{-8pt}

    \caption*{\textbf{\textit{Figure S.3.} Collaboration and citation preferences across research fields.} 
    {\textbf{a}}. Collaboration preference scores across four research fields, comparing countries’ domestic collaboration and their collaboration with the United States and China. {\textbf{b}}. Citation preference scores across the same fields, comparing self-citation and citations directed to the United States and China. Bars represent the mean preference score across countries within each field.
    }
    \label{fig:s_collab_citation_preference}
\end{figure}

\begin{figure}[htbp]
    \centering
    \includegraphics[width=0.9\textwidth]{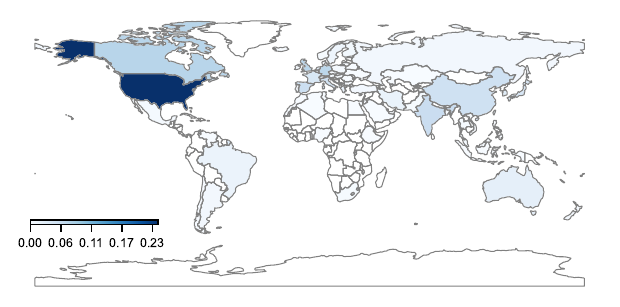}
    \caption{\textbf{a}}
    
    \vspace{2mm} 
    
    \includegraphics[width=0.9\textwidth]{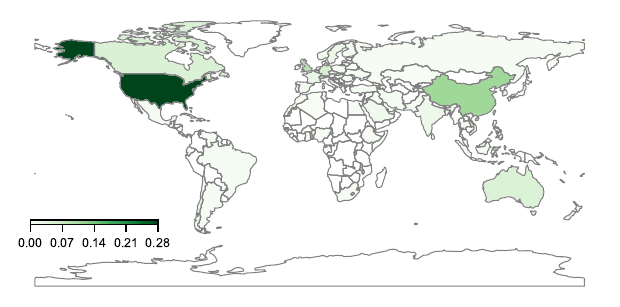}
    \caption{\textbf{b}}

    \caption*{\textbf{\textit{Figure S.4.} Origin and destination countries of migrant scholars.} 
    The map displays the geographic distribution of migrant scholars' countries of origin (\textbf{a}) and destination (\textbf{b}). Color intensity represents the number of migrant scholars associated with each country. Countries with darker shades are either major sources (in the origin map) or major destinations (in the destination map) of international migrant scholars.}
    \label{fig:migration_map}
\end{figure}

\begin{figure}[htbp]
    \centering
    \includegraphics[width=\textwidth]{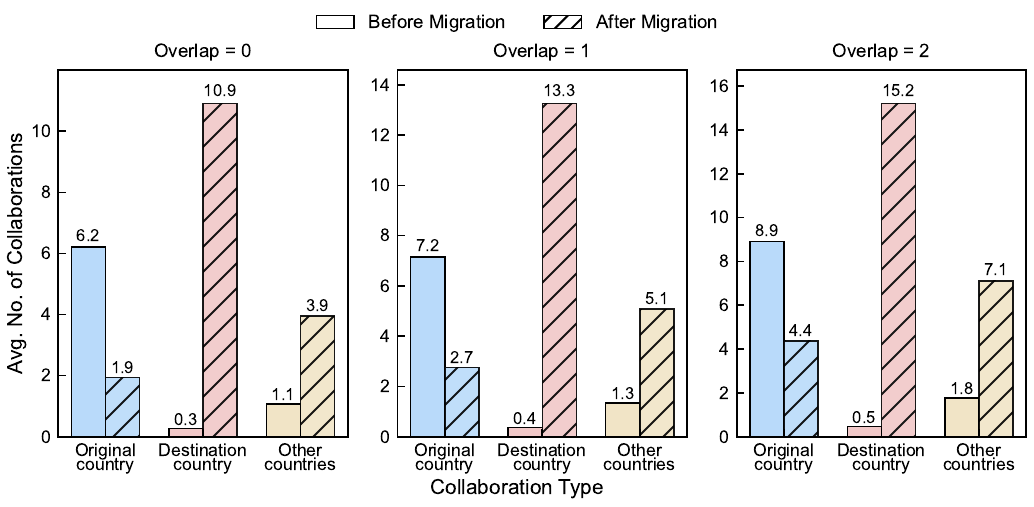}
    \caption*{\textbf{\textit{Figure S.5.} Comparison of international collaboration patterns by overlap (0, 1, 2).} Blank bars indicate collaboration patterns before migration, while hatched bars represent those after migration. Different colors correspond to average number of collaborations with the original country, destination country, and other countries, respectively.}
    
    \label{fig:s5}
\end{figure}

\begin{figure}[htbp]
    \centering
    \includegraphics[width=\textwidth]{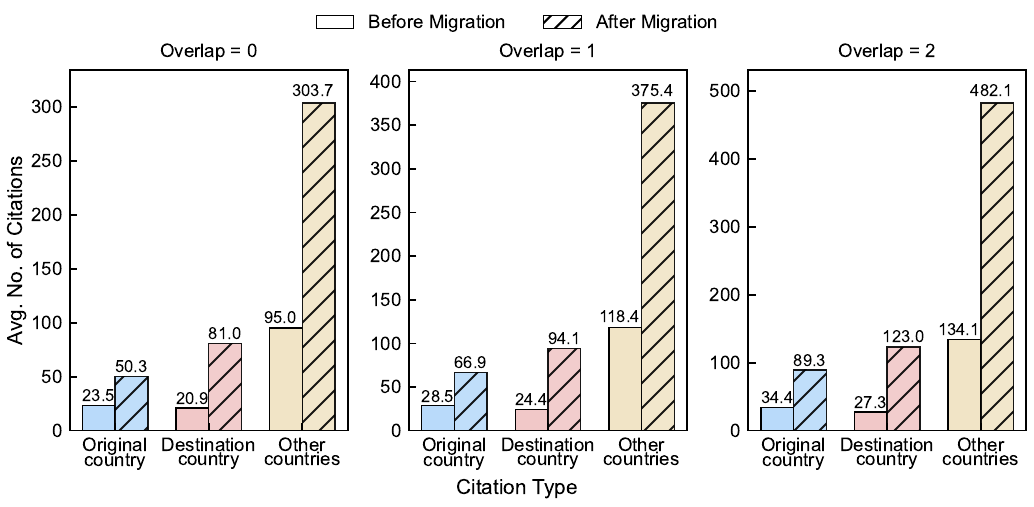}
    \caption*{\textbf{\textit{Figure S.6.} Comparison of international citation patterns by overlap (0, 1, 2).} Blank bars indicate citation patterns before migration, while hatched bars represent those after migration. Different colors correspond to average number of citations with the original country, destination country, and other countries, respectively.}
    \label{fig:s6}
\end{figure}

\begin{figure}[htbp]
    \centering
    \includegraphics[width=\textwidth]{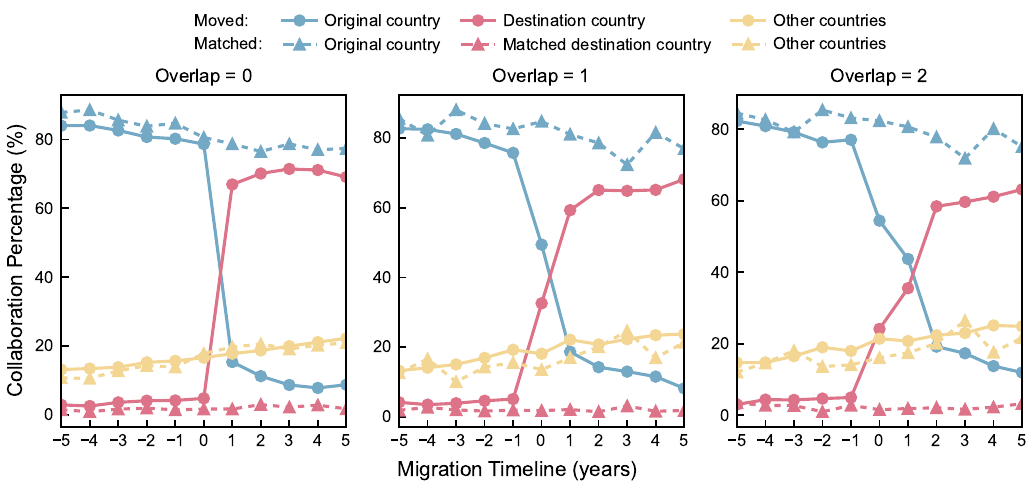}
    \caption*{\textbf{\textit{Figure S.7.} Comparison of collaboration patterns over time by overlap (original side).} Solid lines represent migrant scholars, while dashed lines indicate non-migrant scholars matched on the country of origin.}
    \label{fig:s7}
\end{figure}

\begin{figure}[htbp]
    \centering
    \includegraphics[width=\textwidth]{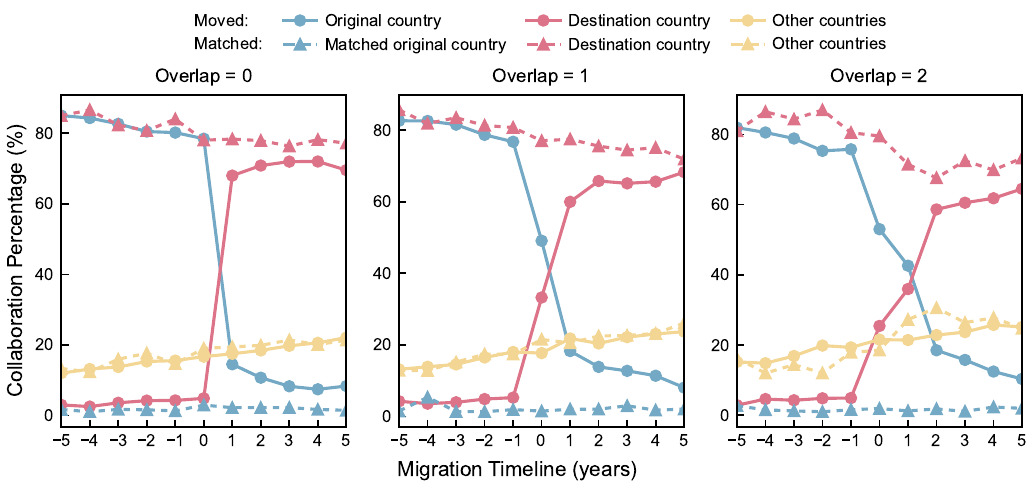}
    \caption*{\textbf{\textit{Figure S.8.} Comparison of collaboration patterns over time by overlap (destination side).} Solid lines represent migrant scholars, while dashed lines indicate non-migrant scholars matched on the country of destination.}
    \label{fig:s8}
\end{figure}

\begin{figure}[htbp]
    \centering
    \includegraphics[width=\textwidth]{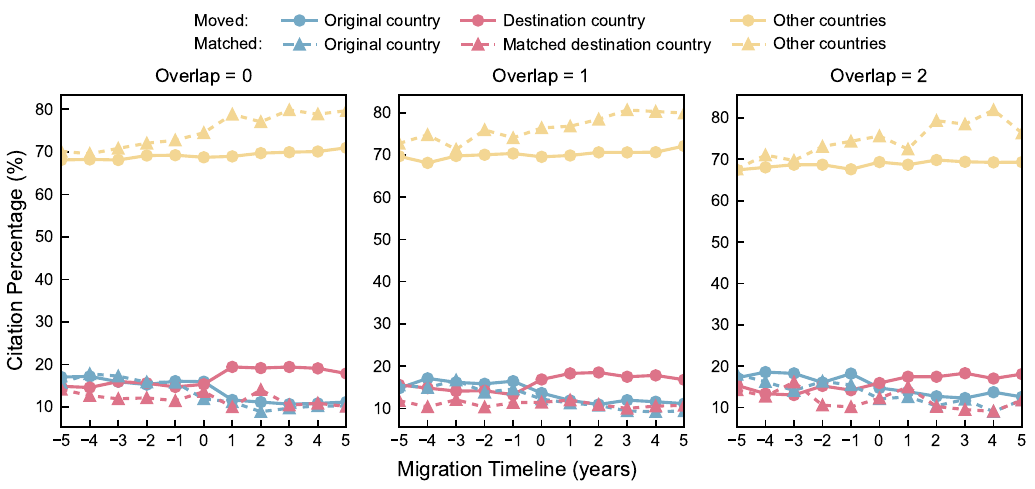}
    \caption*{\textbf{\textit{Figure S.9.} Comparison of citation patterns over time by overlap (original side).} Solid lines represent migrant scholars; dashed lines represent non-migrant scholars from the same country of origin.}
    \label{fig:s9}
\end{figure}

\vspace{0.1cm}

\begin{figure}[htbp]
    \centering
    \includegraphics[width=\textwidth]{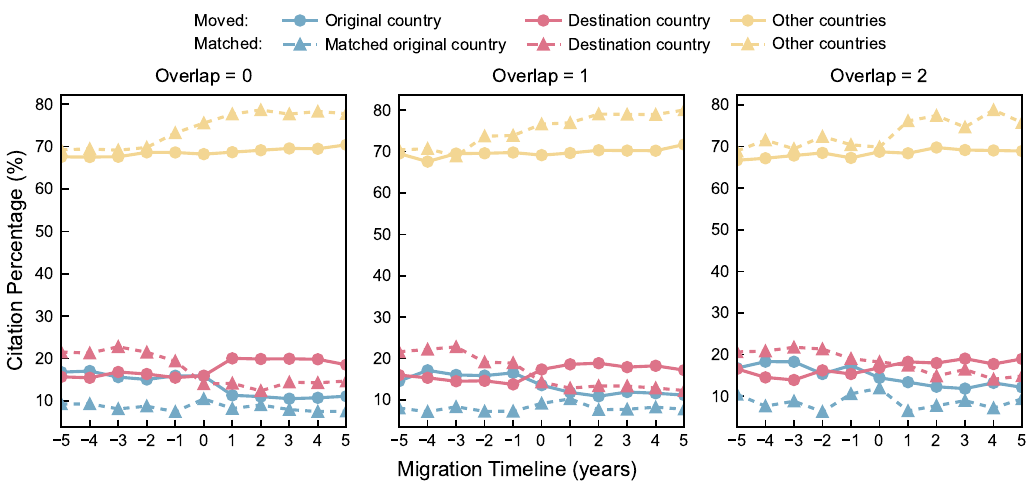}
    \caption*{\textbf{\textit{Figure S.10.} Comparison of citation patterns over time by overlap (destination side).} Solid lines represent migrant scholars; dashed lines represent non-migrant scholars from the same country of destination.}
    \label{fig:s10}
\end{figure}

\FloatBarrier
\newpage


\begin{table}[htbp]
\centering
\captionsetup{width=\textwidth}
\caption*{\textbf{\textit{Table S.1.} Mapping of openAlex concepts to field names.}}
\label{tab:concept_name}
\begin{tabular}{ll}
\toprule
\textbf{Concept Name} & \textbf{Field Name} \\
\midrule
Medicine & Health Sciences \\
Biology & Life Sciences \\
Business & Social Sciences \\
History & Social Sciences \\
Political science & Social Sciences \\
Art & Social Sciences \\
Psychology & Social Sciences \\
Philosophy & Social Sciences \\
Sociology & Social Sciences \\
Economics & Social Sciences \\
Geography & Social Sciences \\
Mathematics & Physical Sciences \\
Materials science & Physical Sciences \\
Computer science & Physical Sciences \\
Chemistry & Physical Sciences \\
Environmental science & Physical Sciences \\
Engineering & Physical Sciences \\
Physics & Physical Sciences \\
Geology & Physical Sciences \\
\bottomrule
\end{tabular}
\end{table}

\vspace{1pt}

\begin{table}[htbp]
\centering
\caption*{\makebox[\linewidth][c]{\textbf{\textit{Table S.2.} Mapping of country codes to country names.}}}
\label{tab:country_codes}
\begin{tabular}{ll}
\toprule
\textbf{Code} & \textbf{Country} \\
\midrule
US & United States \\
CN & China \\
GB & United Kingdom \\
JP & Japan \\
FR & France \\
DE & Germany \\
\bottomrule
\end{tabular}
\end{table}

\begin{table}[htbp]
\centering
\captionsetup{width=\textwidth}
\caption*{\textbf{\textit{Table S.3.} Number of scholars by migration scenario.}}
\label{tab:no_overlap}
\begin{tabular}{lc}
\toprule
\textbf{Overlap Years} & \textbf{No. of Scholars} \\
\midrule
0 & 5091 \\
1 & 2387 \\
2 & 701 \\
\bottomrule
\end{tabular}
\end{table}

\end{document}